\documentclass[prl,footinbib,twocolumn,pacs,notitlepage,amsmath,amssymb,superscriptaddress]{revtex4-2}

\usepackage{amssymb}
\usepackage{amsmath}
\usepackage{comment} 

\usepackage{xr}
\makeatletter
\newcommand*{\addFileDependency}[1]{
  \typeout{(#1)}
  \@addtofilelist{#1}
  \IfFileExists{#1}{}{\typeout{No file #1.}}
}
\makeatother

\newcommand*{\myexternaldocument}[1]{%
    \externaldocument{#1}%
    \addFileDependency{#1.tex}%
    \addFileDependency{#1.aux}%
}

\myexternaldocument{supp}

\usepackage{xcolor}
\usepackage[pdftex]{graphicx}

\usepackage{empheq}

\usepackage{mathtools}

\usepackage{makecell}

\usepackage{siunitx}
\ifdefined\unit\else
  \ifdefined\NewCommandCopy
    \NewCommandCopy\unit\si
  \else
    \NewDocumentCommand\unit{O{}m}{\si[#1]{#2}}
  \fi
\fi

\usepackage{enumerate}
\usepackage{mathtools}
\usepackage{stmaryrd}

\usepackage{enumitem}

\usepackage{textcomp}
\usepackage{gensymb}

\newcommand{\ex}[1]{\mathrm{e}^{#1}}

\newcommand{\dd}[0]{\mathrm{d}}
\newcommand{\ii}[0]{\mathrm{i}}

\newcommand{\EE}[0]{\boldsymbol{E}}

\newcommand{\rr}[0]{\boldsymbol{r}}

\newcommand{\PP}[0]{\boldsymbol{P}}

\newcommand{\qq}[0]{\boldsymbol{q}}
\newcommand{\vv}[0]{\boldsymbol{v}}

\newcommand{\kB}[0]{k_{\mathrm{B}}}

\newcommand{\uu}[0]{\hat{\boldsymbol{u}}}

\newcommand{\rotop}[0]{\boldsymbol{\mathcal{R}}}

\newcommand{\jj}[0]{\boldsymbol{j}}

\definecolor{darkblue}{rgb}{0,0,0.6}
\definecolor{darkred}{rgb}{0.6,0,0}
\definecolor{forestgreen}{rgb}{0.13,0.55,0.13}
\usepackage[colorlinks=true,urlcolor=darkblue,citecolor=darkblue,linkcolor=darkred,linkcolor=forestgreen]{hyperref}

\begin{document}

\title{Stochastic density functional theory for ions in a polar solvent}

\author{Pierre Illien}
\affiliation{Sorbonne Universit\'e, CNRS, Laboratoire PHENIX (Physico-Chimie des Electrolytes et Nanosyst\`emes Interfaciaux), 4 Place Jussieu, 75005 Paris, France}

\author{Antoine Carof}
\affiliation{Université de Lorraine, CNRS, Laboratoire de Physique et Chimie Théoriques, UMR 7019, FR-54000, Nancy, France}

\author{Benjamin Rotenberg}
\affiliation{Sorbonne Universit\'e, CNRS, Laboratoire PHENIX (Physico-Chimie des Electrolytes et Nanosyst\`emes Interfaciaux), 4 Place Jussieu, 75005 Paris, France}
\affiliation{R\'eseau sur le Stockage Electrochimique de l'Energie (RS2E), FR CNRS 3459, 80039 Amiens Cedex, France}

\date{\today}

\begin{abstract}
In recent years, the theoretical description of electrical noise and fluctuation-induced effects in electrolytes has gained a renewed interest, enabled by stochastic field theories like stochastic density functional theory (SDFT). Such models, however, treat solvents implicitly, ignoring their generally polar nature. In the present study, starting from microscopic principles, we derive a fully explicit SDFT theory that applies to ions in a polar solvent. These equations are solved  to compute observables like dynamic charge structure factors and dielectric susceptibilities. We unveil the relative importance of the different contributions (solvent, ions, cross terms) to the dynamics of electrolytes, which are key to understand the couplings between ions and the fluctuations of their microscopic environment.
\end{abstract}

\maketitle

\emph{Introduction.---} Predicting analytically the structure and dynamics of electrolytes from microscopic considerations is a crucial challenge in chemical physics, with significant applications across various fields, from electrochemistry to soft matter physics. Going beyond their well-established average properties, such as their activity coefficient or their conductivity \cite{Harned1943, Robinson2002}, recent experimental research has unveiled the prominent role played by fluctuations on their dynamics \cite{Heerema2015, VASILESCU1974181, Mathwig2012, Secchi2016}. From the theoretical point of view,   particle-based simulations (molecular or Langevin dynamics) as well as  analytical descriptions of the underlying stochastic dynamics have enabled the study of the dynamical response properties of electrolytes, for instance through frequency-dependent conductivity~\cite{Chandra2000a,Yamaguchi2007} or  fluctuation-induced effects~\cite{Dean,Dean2014,Peraud2017,Lee2018, Mahdisoltani2021a,Mahdisoltani2021,{Du2024}}. In particular, stochastic density functional theory (SDFT), that stems from the works of Kawasaki~\cite{Kawasaki1994} and Dean~\cite{Dean1996} on the coarse-graining of interacting Langevin processes, has become a key tool to study the properties of electrolytes (such as their conductivity~\cite{Demery2015a,Peraud2016,Donev2019, Avni2022,Avni2022a,Bonneau2023,Minh2023,Berthoumieux2024}, ionic correlations~\cite{Frusawa2020,Frusawa2022} or viscosity~\cite{Wada2005,Okamoto2022,Robin2024}) as well as the static structure of polar liquids~\cite{Dejardin2018}.

\begin{figure}[b]
\begin{center}
\includegraphics[width=0.44\columnwidth]{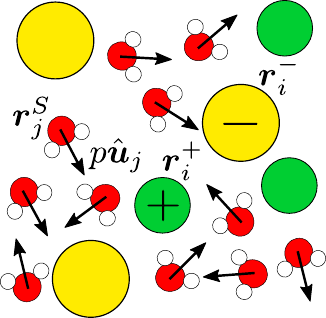}
\includegraphics[width=0.44\columnwidth]{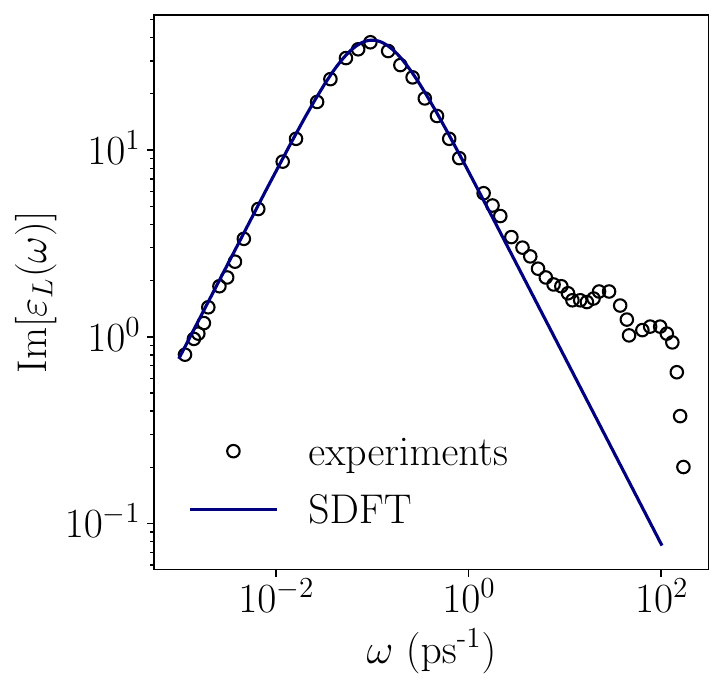}
\caption{Left: System under study: $N_I$ anions and $N_I$ cations, at positions $\rr_i^\pm$, are embedded in a solvent, made of $N_S$ molecules, described by their positions $\rr_j^S$ and their orientations $\uu_j$. Right: Imaginary part of the longitudinal permittivity of pure water, computed theoretically from our SDFT approach, and compared to the experimental data reproduced from Refs.~\cite{Bopp1998,Kaatze1981,Afsar1977,Ruse1971}.}
\label{fig_calibration}
\end{center}
\end{figure}

However, so far, SDFT has been limited to situations where the solvent in which the ions are embedded is described in an implicit way: this approach could therefore be seen as a stochastic extension of the classical Poisson-Nernst-Planck framework.  Although it is able to capture some essential properties of electrolytes, it ignores the fact the solvent molecules also bear charge distribution, which can, to leading order, be represented by a polarization field, whose dynamics is coupled to the ionic charge density. Such couplings are key to understand the relationship between ions and their microscopic environment, which control numerous key processes, such as  water autodissociation \cite{Geissler2001}, dielectric solvation \cite{Song1996}, or NMR relaxation of quadrupolar nuclei \cite{Chubak2021,Chubak2023}. The effect of the solvent on the ion dynamics has been studied from the numerical and theoretical point of view \cite{Hubbard1977,Hubbard1978,Koneshan1998,Banerjee2019a}, and it can be introduced in implicit solvent theories of the conductivity \cite{Chandra2000a}. However, the mutual coupling of ion and solvent dynamics has received less attention. This aspect can be investigated using molecular simulations with an explicit solvent \cite{Sega2013,Sega2014,Minh2023a}, but is much more challenging in analytical approaches.

The goal of this Letter is two-fold. First, we fill the theoretical gap that is set out in this introduction, and derive SDFT equations for ions in a polar solvent: starting from microscopic principles, our approach results in coupled stochastic equations obeyed by the ionic charge density and by the solvent polarization field. Second, these equations are solved to compute the correlation and response functions of the electrolytes, unveiling the relative importance of the different contributions (solvent, ions, cross terms) to the dynamics. As a perspective, given the strong interest raised by the use of SDFT to study charged systems at the level of fluctuations, we expect that the present work will bring insight in the fluctuation-induced effects that have been  recently evidenced in electrolytes ~\cite{Dean,Dean2014,Peraud2017,Lee2018, Mahdisoltani2021a,Mahdisoltani2021,{Du2024}}.

\emph{Model.---} We consider a binary electrolyte made of $N_I$ cations and $N_I$ anions of respective charge $\pm z e$. We denote by $\rr_i^\pm(t)$ their positions. 
The electrolyte is embedded in a solvent represented by $N_S$ point charge dipoles, of dipolar moment $p$. 
We denote by $\rr^S_j(t)$ their position and $\uu_j(t)$ their unit orientation vector (see Fig.~\ref{fig_calibration}). Even though some of the previous SDFT approaches accounted for hydrodynamic couplings \cite{Avni2022,Avni2022a,Peraud2016,Peraud2017}, the goal of the present study is to highlight the interplay between ionic current and solvent polarization dynamics. To this end, we will adopt a simple description, where both ions and solvent molecules obey overdamped Langevin dynamics. We denote by $\mu_I$ (resp. $\mu_S$) the bare mobility of the ions (resp. solvent molecules), and by $D_I=\kB T \mu_I$ (resp. $D_S=\kB T \mu_S$) the associated bare diffusion coefficients.  We also denote by $D_S^r$ is the rotational diffusion coefficient of the solvent molecules, and $\mu_S^r$ the associated orientational mobility ($D_S^r=\kB T \mu_S^r$). We introduce the stochastic density of cations and anions as $n_\pm(\rr,t)  =\sum_{i=1}^{N_I} \delta(\rr-\rr_i^\pm(t))$, and that of solvent molecules, which has both translational and orientational degrees of freedom: $n_S(\rr,\uu,t) = \sum_{j=1}^{N_S} \delta(\rr-\rr_j^S(t))\delta(\uu-\uu_j(t))$.

The electrostatic potential at a given point of the solution $\rr$, denoted by $\varphi(\rr)$, is obtained by solving Poisson's equation: $-\nabla^2\varphi(\rr)=[\rho_I(\rr)+\rho_S(\rr)]/\varepsilon_0$ \footnote{Note that we make the choice to use SI units throughout the paper: the mapping to cgs or Gaussian units, which are still used by many authors, is typically done through the change $\varepsilon_0 \to 1/(4\pi)$.}, where $\rho_I=ze(n_+-n_-)$ is the charge density of the ions and $\rho_S$ is the charge density associated to the solvent. Within the dipolar approximation (i.e. at distances much larger than the typical size of a solvent molecule), $\rho_S(\rr) \simeq -\nabla\cdot \PP(\rr)$, where we define the polarization field of the solvent as $\PP(\rr,t) =  \int \dd \uu \; p \uu n_S(\rr,\uu, t)$.   The solution to Poisson's equation is then $\varphi(\rr,t) = \int \dd \rr' \; \mathcal{G}(\rr-\rr') [\rho_I(\rr',t)-\nabla\cdot \PP(\rr',t)]$, with the Green function $ \mathcal{G}(\rr)=1/(4\pi\varepsilon_0 |\rr|)$.

\emph{Stochastic density functional theory.---} In the overdamped limit, the state of the system is fully characterized by the positions of the ions $\rr_i^\pm$, and the position and orientations of the solvent molecules $\rr_j^S$ and $\uu_j$. The evolution equations read:
\begin{align}
\frac{\dd \rr_i^\pm}{\dd t} &=\mp \mu_I ze\nabla\varphi(\rr_i^\pm) + \sqrt{2D_\pm} \boldsymbol{\xi}_i^\pm(t), \label{langevin_rions}\\
\frac{\dd\rr^S_j}{\dd t} &=-\mu_S(p \uu_j\cdot \nabla) \nabla\varphi(\rr_j^S) + \sqrt{2D_S} \boldsymbol{\xi}_j^S(t), \label{langevin_rs}\\
\frac{\dd \uu_j}{\dd t} &= \left\{-[\mu_S^r p\uu_j \times \nabla\varphi(\rr_j^S)] + \sqrt{2D_S^r} \boldsymbol{\xi}_j^{S,r}(t) \right\}\times \uu_j, \label{langevin_ua}
\end{align}
where the noises $\boldsymbol{\xi}_a^\alpha(t)$ are uncorrelated Gaussian white noises of zero average and unit variance, i.e. $\langle \xi_{i,n}^\alpha (t)\xi_{j,m}^\beta(t') \rangle = \delta_{\alpha\beta} \delta_{mn} \delta_{ij} \delta(t-t')$, where $n$ or $m$ are components of the vectors. Note that we assumed here that the ions only interact through electrostatic interactions, and that short-range repulsive interactions are neglected.

Starting from Eqs.~\eqref{langevin_rions}-\eqref{langevin_ua}, the derivations of the  equations satisfied by the ionic charge density $\rho_I=ze(n_+-n_-)$ and the total number density $\mathcal{C} = n_++ n_-$ of ions are standard and rely on It\^o's lemma~\cite{Dean1996,Demery2015a}. One gets:
\begin{eqnarray}
    \partial_t \rho_I &=& D_I\nabla^2 \rho_I+\mu_I (ze)^2 \nabla \cdot (\mathcal{C} \nabla \varphi) \nonumber\\
    &&+\nabla \cdot (ze\sqrt{2D_I\mathcal{C}} \boldsymbol{\zeta}_{\rho_I}), \label{DK1}\\
    \partial_t \mathcal{C} &=& D_I\nabla^2 \mathcal{C}+\mu_I  \nabla \cdot (\rho_I \nabla \varphi)+\nabla \cdot (\sqrt{2D_I\mathcal{C}} \boldsymbol{\zeta}_\mathcal{C}), \label{DK2}
\end{eqnarray}
where $\boldsymbol{\zeta}_\alpha(\rr,t)$ are uncorrelated Gaussian white noises of zero average and unit variance (and similarly in the next equation). In contrast, the derivation of the equation satisfied by $n_S$ is more subtle as it involves two variables, one of them being an orientational degree of freedom, see Supplementary Material (SM)~\footnote{See Supplemental Material, which includes Refs. \cite{Gardiner1985,
		Dean1996,
		Ahmadi2006,
		Brotto2013,
		Golestanian2019,
		Berthoumieux2015,
		Berthoumieux2018,
		Vatin2021,
		Maggs2006,
		Blossey2022,
		Blossey2022a,
		Becker2023,
		Hansen1986,
		Ladanyi1999,
		Caillol1987,
		Madden1984,
		Caillol1986,
		Nakamura2009,
		Demery2015,
		Das2013}, for additional information about the analytical calculations.}. It reads:
\begin{align}
&        \partial_t n_S(\rr,\uu,t) =  D_S\nabla^2_{\rr}n_S + D_S^r \rotop_{\uu}^2 n_S \label{DK3}\\
        &  +\nabla_{\rr} \cdot [n_S \mu_S p (\uu\cdot \nabla) \nabla\varphi(\rr)] +  \rotop_{\uu}\cdot [n_S\mu_S^r p \uu \times \nabla \varphi(\rr)] \nonumber\\
    &+\nabla_{\rr} \cdot(\sqrt{2D_S n_S } \boldsymbol{\zeta}_S(\rr,\uu,t)) \nonumber\\
&    +\rotop_{\uu} \cdot(\sqrt{2D_S^r n_S } \boldsymbol{\zeta}_{S,r}(\rr,\uu,t))  \nonumber
\end{align}
where $  \rotop_{\uu} = \uu \times \nabla_{\uu}$ is the rotational gradient operator~\cite{Doi1988}.  Finally, note that, within the dipolar approximation, Eqs.~\eqref{DK1}-\eqref{DK3} form a closed set of equations for $\rho_I$, $\mathcal{C}$ and $n_S$, since the electrostatic potential $\varphi$ is related to $\rho_I$ and $\rho_S$ through $\varphi(\rr,t) = \int \dd \rr' \; \mathcal{G}(\rr-\rr') [\rho_I(\rr',t)+\rho_S(\rr',t)]$.

Although Eqs.~\eqref{DK1}-\eqref{DK3} are exact reformulations of the microscopic overdamped Langevin dynamics, they are rather unpractical, since they are nonlinear. To make analytical progress, we linearize them around a uniform, time-independent state~\cite{Dean2014,Demery2014}. More precisely, denoting by $C_I$ the concentration of the electrolyte, we write $n_\pm = C_I + \delta n_\pm$, which implies $\rho_I=ze(\delta n_+-\delta n_-)$. At linear order in the perturbation, the equation for $\rho_I$ is decoupled from that on $c$:
\begin{equation}
 \partial_t \rho_I(\rr,t) = D\nabla^2 \rho_I+2\mu (ze)^2 C_I \nabla^2\varphi+ze\sqrt{4DC_I} {\eta}_{\rho_I}.\label{lineq_rho}  
\end{equation}
where $\langle \eta_{\rho_I}(\rr,t) \eta_{\rho_I}(\rr',t')\rangle=-\delta(t-t') \nabla_{\rr}^2\delta(\rr-\rr')$. Similarly, the density of the solvent is expanded as $n_S = {C_S}/{4\pi} + c_S$, where $C_S$ is the concentration of solvent molecules, and the $1/4\pi$ factor accounts for the uniform density of orientations. The equation satisfied by the polarisation is deduced using $\PP(\rr,t) =  \int \dd \uu \; p \uu c_S(\rr,\uu, t)$:
\begin{align}
&    \partial_t \PP(\rr,t) = D_S \nabla^2 \PP -2D_S^r \PP \nonumber\\
    &+\frac{1}{3}p^2 C_S \nabla(\mu_S \nabla^2 \varphi-2\mu_S^r \varphi) + \boldsymbol{\Xi} (\rr,t),
    \label{lineq_P2}
\end{align}
with $\langle \Xi_k(\rr,t) \Xi_l(\rr',t') \rangle= \frac{2 p^2 C_S }{3} \delta_{kl} \delta (t-t') (-D_S \nabla^2+2D_S^R) \delta (\rr-\rr')$.  Note that, for a system without ions, the deterministic term in rhs of Eq.~\eqref{lineq_P2} can be rewritten under the form $-\frac{\delta}{\delta \PP} \mathcal{F}[\PP]$, where the free energy functional $\mathcal{F}[\PP]$ is here derived from microscopic considerations~(see SM), and may be compared to previously proposed expressions based on symmetry considerations~\cite{Maggs2006,Berthoumieux2015,Berthoumieux2018,Blossey2022,Blossey2022a,Becker2023}.

\emph{Evolution equations for the charge densities.---} Finally, the evolution equation for $\rho_S$ is deduced by taking the divergence of Eq.~\eqref{lineq_P2}. In Fourier space \footnote{Throughout the paper, we will use the following convention for spatial Fourier transforms: $\tilde{f}(\qq) = \int \dd \rr \; \ex{-\ii \qq\cdot \rr} f(\rr)$, and $f(\rr,t) = \frac{1}{(2\pi)^3}\int \dd \qq \; \ex{\ii \qq\cdot \rr}  \tilde{f}(\qq)$},  we find the following coupled stochastic  equations for the ion and solvent charge densities:
\begin{equation}
    \partial_t \begin{pmatrix}
        \tilde \rho_I(\qq,t) \\
        \tilde \rho_S(\qq,t)
    \end{pmatrix}
=- \mathbf{M}
\begin{pmatrix}
        \tilde \rho_I(\qq,t) \\
        \tilde \rho_S(\qq,t)
    \end{pmatrix}
    +
    \begin{pmatrix}
        \Xi_I(\qq,t) \\
        \Xi_S (\qq,t)
    \end{pmatrix},
    \label{eq_main_result}
\end{equation}
with
\begin{equation}
\label{ }
\mathbf{M} = 
\begin{pmatrix}
  D_I(q^2+\kappa_I^2)    &  D_I \kappa_I^2  \\
    D_S \kappa_S(q)^2   &    D_S(q^2+\kappa_S(q)^2) +2D_S^r
\end{pmatrix},
\end{equation}
where we introduced the Debye screening length of the electrolyte  $\kappa_I^{-1}$, which is such that $\kappa_I^2 = \frac{2C_I (ze)^2}{\varepsilon_0 \kB T}$,  and the wavenumber-dependent screening length associated with the charge polarization of the solvent $\kappa_S (q)^2  =\frac{2p^2 C_S }{3\varepsilon_0  \kB T a^2} \left(1+ \frac{q^2a^2}{2}\right) $, where we introduced $a\equiv(D_S/D_S^r)^{1/2}$, which is the typical size of a solvent molecule. The noises $\Xi_I$ and $\Xi_S$ have zero average and correlations $\langle \Xi_\alpha (\qq,t) \Xi_\beta(\qq',t') \rangle = 2q^2 \varepsilon_0  \kB T D_\alpha \kappa_\alpha^2    (2\pi)^d \delta (\qq+\qq') \delta(t-t') \delta_{\alpha\beta} $. Eq.~\eqref{eq_main_result} is formally solved as $\tilde\rho_\alpha(\qq,t) = \int_{-\infty}^t \dd t'\; \mathcal{M}_{\alpha\beta}(t-t') \Xi_\beta(\qq,t')$, where $ \boldsymbol{\mathcal{M}}(t)=\exp(-t \mathbf{M})$, and where repeated indices are implicitly summed upon.

Eq.~\eqref{eq_main_result} is the central result of this Letter, and several comments are in order: (i) The coupled equations for the fields $\tilde\rho_I$ and $\tilde\rho_S$  are fully explicit and actually depend on a reduced set of parameters: the self-diffusion coefficients of ions and solvent molecules $D_I$ and $D_S$, as well as their respective screening lengths $\kappa_I$ and $\kappa_S(q)$, which are themselves known in terms of the microscopic parameters of the model. They do not rely on prior knowledge of the static structure of the electrolyte, whereas it is generally the case in dynamical DFT approaches, and are expected to be valid for dilute electrolytes; (ii) These equations give access to the coupled statistics of the ionic charge density $\rho_I$ and to the charge polarization $\rho_S$, as well as to many derived quantities of experimental interest, that will be described in what follows; (iii) It is straighforward to show that the eigenvalues of $\mathbf{M}$ are always positive, in such a way that the solutions to Eq.~\eqref{eq_main_result} never diverge (which is a consequence of the mutual relaxation of ionic currents and solvent polarization); (iv) As a consequence of the linearization procedure, the stochastic fields $\tilde\rho_I$ and $\tilde\rho_S$ have Gaussian statistics.

\emph{Static and dynamic charge structure factors.---}  From the solution to the matrix equation obeyed by the charge density fields $\tilde \rho_I$ and $\tilde \rho_S$, we deduce the charge dynamic structure factors (or intermediate scattering functions), which are defined as $F_{\alpha\beta}(q,t) = (\mathcal{N}_\alpha\mathcal{N}_\beta)^{-1/2} \langle \tilde \rho_\alpha(\qq,t) \tilde\rho_\beta(-\qq,0)  \rangle $~\cite{Hansen1986}, where we write $\mathcal{N}_I=2 N_I$ and $\mathcal{N}_S= N_S$ for simplicity (we also write $C^\text{tot}_I=2C_I$ and $C^\text{tot}_S=C_S$). Note that this definition actually holds for a finite system in a volume $V$: our results, which were implicitly derived in the thermodynamic limit, can be mapped onto finite systems, which are of interest in molecular dynamics simulations for instance, by making the change $(2\pi)^d \delta(\qq+\qq') \to V \delta_{\qq,-\qq'}$. We find:
\begin{equation}
F_{\alpha\beta}(q,t)= \frac{ 2q^2\varepsilon_0 \kB T }{(C^\text{tot}_\alpha C^\text{tot}_\beta)^{1/2}}  \sum_{\substack{\gamma\in\{S,I\}   \\ \nu =\pm1   }} D_\gamma \kappa_\gamma^2 C_{\alpha \gamma,\beta\gamma}^{(\nu)} \ex{-\lambda_\nu |t|},
\label{ISF_full}
\end{equation}
where the coefficients of the fourth-rank tensors $\mathbf{C}^{(\nu)}$ can be expressed simply in terms of the eigenvalues $\lambda_\nu$ and entries of $\mathbf{M}$ (see SM). From this expression, we also define the static charge structure factors $\mathcal{S}_{\alpha\beta}(q)=F_{\alpha\beta}(q,0)$ as well as the dynamic charge structure factors in frequency space: $\tilde F_{\alpha\beta}(q,\omega) \equiv \int \dd t \; \ex{-\ii\omega t} F_{\alpha\beta}(q,t)$.

The total structure factor is $F_\text{tot}(q,t) = \sum_{\alpha,\beta\in\{ I,S\}}  \frac{(C^\text{tot}_\alpha C^\text{tot}_\beta)^{1/2}}{C^\text{tot}_I + C^\text{tot}_S} F_{\alpha\beta}(q,t)$. We can verify that our model fulfils the electroneutrality condition, as $\lim_{q\rightarrow 0} F_\text{tot}(q,t) = 0$ for any $t$. In particular, the static total structure factor follows the second moment Stillinger-Lovett condition, $\mathcal{S}_\text{tot}(q) = F_\text{tot}(q,0) \sim \varepsilon_0 \kB T q^2$, which reflects the fact that our system is a perfect conductor \cite{Martin1983}. In addition, in the static limit, we get the ion-ion charge structure factor $\mathcal{S}_{II}(q)\sim_{q\to 0} (ze)^2 q^2 \varepsilon_\text{w}/\kappa_I^2$, with $\varepsilon_\text{w}$ the permittivity of pure solvent in this model, which will be discussed below.  This expression coincides with the one that is obtained by considering ions in an implicit solvent of permittivity $\varepsilon_\text{w}$, where $F_\text{imp}(q,t) = \frac{(ze)^2 q^2}{q^2+\kappa_\text{imp}^2} \ex{-D_I(q^2+\kappa_\text{imp}^2 )t}$ (with $\kappa_\mathrm{imp}^2 = \frac{2C_I (ze)^2}{\varepsilon_\text{w}\varepsilon_0 \kB T}$, see SM), from which we deduce $\mathcal{S}^\text{imp}_{II}(q)\sim_{q\to 0} (ze)^2 q^2 /\kappa_\text{imp}^2$.

Finally, to characterize the dynamics of the ions, we define the ionic relaxation time as $\tau_{II}(q) = \tilde F_{II}(q,\omega=0)/2F_{II}(q, t=0)$. In the zero-$q$ limit, the ionic relaxation time of the implicit model is the Debye time, $\tau_{II}^{\text{imp}}(q=0) = \frac{1}{D_I\kappa_\text{imp}^2}$,  while our models gives $\tau_{II}(q=0) =\frac{1}{D_I\kappa_\text{imp}^2} + \frac{ \varepsilon_\text{w} - 1}{2\varepsilon_\text{w}D_S^r}$.  This shows that explicit solvent molecules tend to increase the ionic relaxation time. Note that the Debye model is retrieved when $D_S^r \ll (\varepsilon_\text{w}-1) D_I \kappa_\text{imp}^2/2\varepsilon_\text{w} $, i.e. in the limit where  solvent molecules relax much faster  than ions.

\emph{Dieletric susceptibility.---} Within linear response, if the electrolyte is submitted to a small external field $\EE_0$,  the polarisation of the electrolyte reads, in Fourier space: $\PP(\qq,\omega)=\boldsymbol{\chi}_{\PP}(\qq,\omega) \cdot \EE_0(\qq,\omega)$, and the ionic current, defined as $\jj (\rr,t) = \sum_{\nu=\pm 1}\sum_{i} (\nu ze) \vv_i^\nu(t) \delta(\rr-\rr_i^\nu(t)) $ (where $\vv_i^\nu$ is the velocity of the $i$-th ion of species $\nu$), is related to the external field through $\jj(\qq,\omega) = \boldsymbol{\chi}_{\jj}(\qq,\omega)\cdot \EE_0(\qq,\omega)$, where $\boldsymbol{\chi}_{\PP}$ and $\boldsymbol{\chi}_{\jj}$ are susceptibility tensors, characterizing the response of the solvent and of the ions, respectively. The total susceptibility is $\boldsymbol{\chi} = \boldsymbol{\chi}_{\PP} + \frac{\ii}{\omega \varepsilon_0} \boldsymbol{\chi}_{\jj}$~\cite{Caillol1986}. Computing the susceptibility tensor from microscopic and stochastic quantities is a subtle task, that has been discussed by many authors, both  in the case of pure dipolar solvent~\cite{Madden1984,Bopp1998, Ladanyi1999}, and in the case of an electrolyte~\cite{Caillol1987,Levesque1990,Cox2019}. Here, we will only be interested in the $|\qq| \to 0$ limit of the susceptibility tensor, in which one gets~\cite{Caillol1987}: $\boldsymbol{\chi}_{\PP}(\omega)=(\varepsilon_0\kB T V)^{-1}[\langle \PP \PP \rangle(0)+\int_0^\infty \dd t\; \ex{\ii \omega t} (\ii \omega\langle \PP \PP \rangle(t) + \langle \PP \jj \rangle(t) )]$ and $\boldsymbol{\chi}_{\jj}(\omega)=(\kB T V)^{-1}\int_0^\infty \dd t\; \ex{\ii \omega t} (\langle \jj \jj \rangle(t) +\ii\omega \langle \PP \jj \rangle(t) )$ where we used the shorthand notation $\langle \boldsymbol{A} \boldsymbol{B} \rangle(t) = \lim_{\qq\to 0\\
} \langle  \boldsymbol{A} (\qq,t) \cdot    \boldsymbol{B} (-\qq,0) \rangle $. Importantly, all the correlation functions that appear in these relations can be computed directly from our SDFT approach (see SM).

\emph{Calibration of the model.---} To calibrate the three independent parameters that describe our explicit solvent ($\kappa_S(0)$, $D_S^r$ and $a$), we compute the longitudinal permittivity of pure water, i.e. in the limit $C_I \to 0$ of our model. It is related to the susceptibility defined above through $\varepsilon_L(\omega) = 1/[1-\chi_L(\omega)]$~\cite{Hansen1986}, where $\chi_L$ is the longitudinal part of the susceptibility tensor~\footnote{The longitudinal permittivity is the one measured in experiments. Note that the transverse permittivity can be computed through the relation $\varepsilon_T(\omega) = 1+\chi_T(\omega)$, and has a characteristic time $\tau_S/\varepsilon_\text{w}$, as expected \cite{Gekle2012}}. We get $\varepsilon_L(\omega) = \varepsilon_0 + \frac{\varepsilon_\text{w} -\varepsilon_0}{1-\ii\omega\tau_S} $, with $\tau_S=1/2 D_S^r $ and $\varepsilon_\text{w} = 1+\frac{p^2 C_S }{3\varepsilon_0  \kB T}$. 
This Debye-like dependence of the permittivity on $\omega$ is to be expected, since the polarization relaxes through a linear damping term (see Eq. \eqref{lineq_P2}) at our level of approximation.
Importantly, this provides an estimate of the relative permittivity involved in the implicit description from microscopic considerations, and corresponds to the expression for a gas of interacting dipoles in the mean-field limit \cite{Abrashkin2007}. Comparison with experimental data [Fig.~\ref{fig_calibration}, right] yields $\varepsilon_\text{w} \simeq 78.5\varepsilon_0$ and  $\tau\simeq 10$ ps, which enforces $\kappa_{S}(0) \simeq 12.4/a$ and $D_S^r \simeq 0.05$ ps$^{-1}$. The value of $a$ is chosen in such a way that the (self) translational diffusion coefficient of water molecules $D_S=a^2 D_S^r$ matches the values typically measured at 25\degree C (\num{2.3e-9} \unit{m^2.s^{-1}}): this imposes $a\simeq 2.14$~\AA. Consequently, given the expression of $\kappa_S(0)$, the product $p^2 C_S$ is also fixed: if one imposes $C_S \simeq 55$ M (the typical molar concentration of water under normal conditions), then one finds $p=\num{1.4}~\unit{D}$, which is reasonably close the value of the dipolar moment of a water molecule ($\simeq\num{1.8}~\unit{D}$ in gas phase or $\simeq \num{2.6}~\unit{D}$ in liquid phase \cite{Silvestrelli1999}).

\begin{figure}
\begin{center}
\includegraphics[width=0.49\columnwidth]{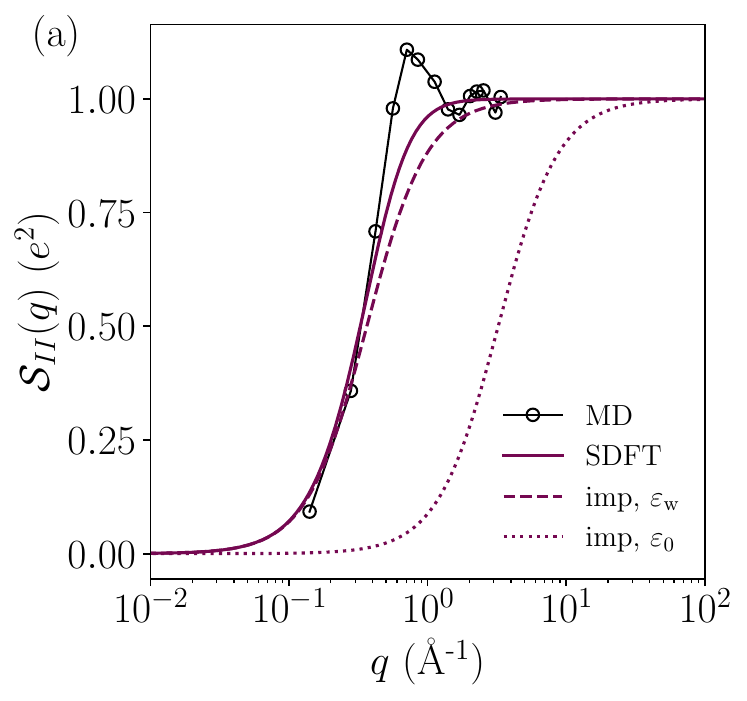}
\includegraphics[width=0.49\columnwidth]{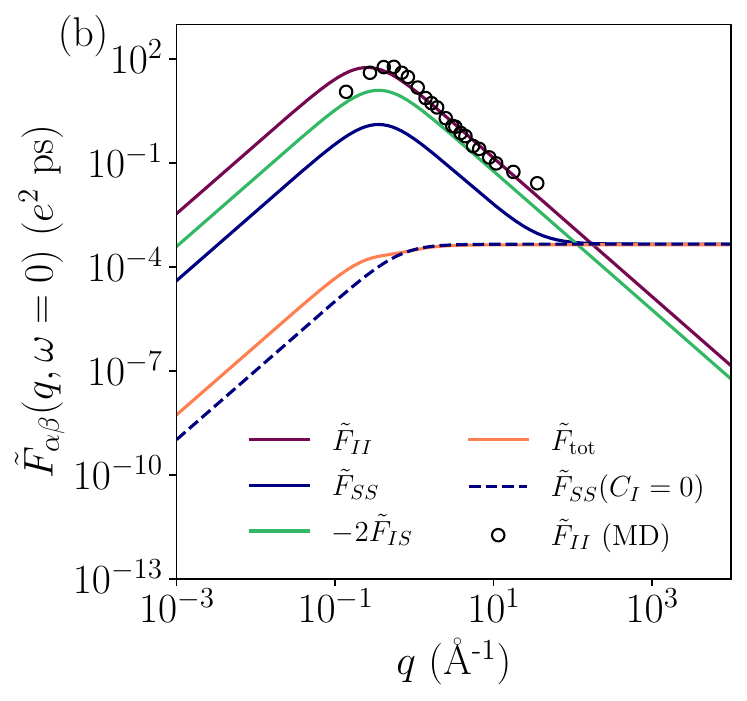}
\includegraphics[width=0.49\columnwidth]{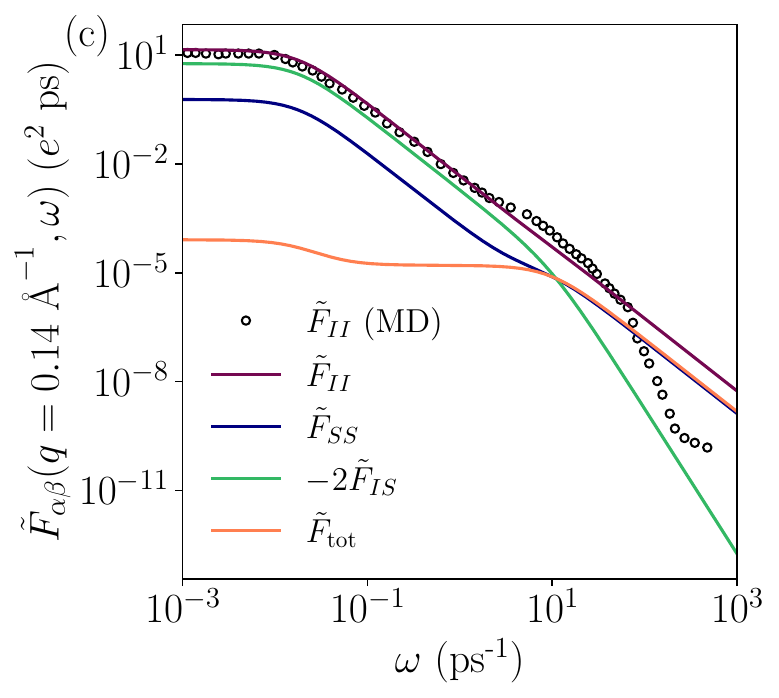}
\includegraphics[width=0.49\columnwidth]{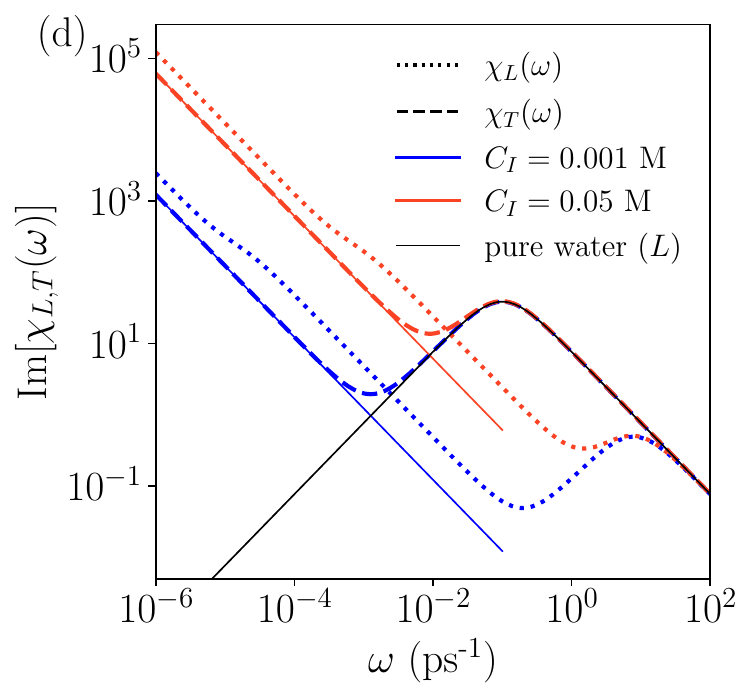}
\caption{
(a)~Static charge structure factor of the ions only, as a function of the wavevector $q$.  Results from our SDFT approach are compared and with the implicit solvent model with permittivity $\varepsilon_\text{w}$ (resp. $\varepsilon_0$), i.e. with Debye length $\kappa_\text{imp}$ (resp. $\kappa_I$).
(b)~Zero-frequency dynamic charge structure factors, where the different contributions are split. For comparison, the result for pure water ($C_I=0$) is shown. 
(c)~Frequency dependence of the dynamic charge structure factors for $q=0.14$~\AA$^{-1}$.
(a-c) In these plots, we consider a 1:1 electrolyte of concentration $C_I=1.2$~M. 
Symbols: results from molecular dynamics simulations reproduced from Ref.~\cite{Minh2023a}.
(d) Imaginary part of the transverse and longitudinal components of the total susceptibility tensor $\boldsymbol{\chi}(\omega)$. Dashed lines: $ \sigma_0/\varepsilon_0\omega$. See SM for other numerical parameters.
}
\label{fig_struct_factors}
\end{center}
\end{figure}

\emph{Numerical estimates.---} Using these numerical parameters, we first compute the static structure factor of the ions $\mathcal{S}_{II}(q)$, and plot it on Fig.~\ref{fig_struct_factors}(a). We find that the expression computed within our explicit solvent model is very close to the expression that can be computed with the implicit model approach, i.e. with the Debye screening length $\kappa_\text{imp}$: this validates the model beyond the $q=0$ limit that was employed for calibration.  We also study the $q$-dependence of the dynamic charge structure factors [Fig.~\ref{fig_struct_factors}(b)] and split the different contributions (ions, solvent, cross terms). Importantly, and just like in molecular dynamics simulations~\cite{Minh2023a}, we observe that the cross water-ion correlations, which are negative, are sufficiently strong to cancel the contributions from ions and solvent at small $q$. At large wavevectors, the total dynamic structure factor is equal to that of pure water, which tends to a constant, as expected in the present case where short-range interactions between molecules are ignored. For a fixed value of the wavevector ($q=0.14$~\AA$^{-1}$, which corresponds to distances where short-range interactions do not play a significant role, and where our theory should be most valid), we plot the frequency-dependence of the dynamic structure factor and obtain a good agreement with results from MD simulations for the ionic contribution [Fig.~\ref{fig_struct_factors}(c)]: this validates our approach beyond the static limit. The analysis of the other contributions show that, at large frequencies, the cross contributions become subdominant, and do not contribute anymore to the total dynamic structure factor.

Finally, we turn to the dielectric response of the electrolyte. The imaginary parts of transverse and longitudinal components of the susceptibility tensor $\boldsymbol{\chi}(\omega)$ are shown on Fig.~\ref{fig_struct_factors}(d). The transverse components display crossovers at the characteristic frequencies $\sim D_I\kappa_\text{imp}^2$ and $\sim 
D_S^r$, which are respectively the Debye relaxation time of the ions and the typical persistence of the orientation of the water molecules. This is consistent with previous observations from molecular simulations~\cite{Sega2013,Sega2014}. Finally, we retrieve that the small frequency limit of the imaginary part of the susceptibility diverges because of the overall conductivity of the system, which is non-zero in the presence of ions. Indeed, one gets $\text{Im}[{\chi}_\alpha(\omega)] \sim  c_\alpha \sigma_0/\varepsilon_0\omega$,
where $\sigma_0 = \varepsilon_0 D_I \kappa_I^2$ is the ideal conductivity of the ions in vacuum, and where $c_\alpha=2$ for $\alpha=L$ and $c_\alpha=1$ for $\alpha=T$.

\emph{Perspectives.---} The present framework opens new directions in the analytical description of experiments such as NMR relaxation of quadrupolar ions \cite{Hynes1981,Perng1998} and ultrafast spectroscopy probing solvation dynamics \cite{Maroncelli1988,Jimenez1994,Song1996,Geissler2000, Roy2015, Banerjee2019a} in electrolyte solutions. In addition, several directions could be followed in order to refine the present theoretical framework: (i) the  short-range interactions between the ions could be accounted for by combining SDFT with closure schemes such as the Mean Spherical Approximation \cite{Dufreche2005,Bernard2023}. Although it is a challenging and longer-term perspective, this could extend the relevance of our model beyond the small-$q$ limit, in particular to compare the $q$-dependence of the solvent permittivity with alternative approaches \cite{Bopp1996,Bopp1998}; (ii) hydrodynamic couplings between the ions and the solvent could be incorporated, in order to ensure momentum conservation within the solvent, as done previously at the SDFT level without taking into account the explicit polarization \cite{Avni2022,Avni2022a,Peraud2016,Peraud2017};  (iii) accounting for reactions, such as ion pairing or water autodissociation, could be an interesting theoretical avenue to explore, building on recent developments of SDFT \cite{Spinney2024}; (iv) finally, going beyond the mean-field coupling between the ionic and solvent densities could be of interest to compute dielectric decrements \cite{Levy2012,Adar2018} and solvent relaxation in the vicinity of ions.

\emph{Acknowledgments.---} The authors thank Michiel Sprik, Steve Cox, Vincent D\'emery, Th\^e Hoang Ngoc Minh, Haim Diamant, and David Andelman for discussions on this topic, and Marie Jardat for comments on the manuscript. This project received funding from the European Research Council under the European Union’s Horizon 2020 research and innovation program (grant agreement no. 863473).


\begin{thebibliography}{89}%
	\makeatletter
	\providecommand \@ifxundefined [1]{%
		\@ifx{#1\undefined}
	}%
	\providecommand \@ifnum [1]{%
		\ifnum #1\expandafter \@firstoftwo
		\else \expandafter \@secondoftwo
		\fi
	}%
	\providecommand \@ifx [1]{%
		\ifx #1\expandafter \@firstoftwo
		\else \expandafter \@secondoftwo
		\fi
	}%
	\providecommand \natexlab [1]{#1}%
	\providecommand \enquote  [1]{``#1''}%
	\providecommand \bibnamefont  [1]{#1}%
	\providecommand \bibfnamefont [1]{#1}%
	\providecommand \citenamefont [1]{#1}%
	\providecommand \href@noop [0]{\@secondoftwo}%
	\providecommand \href [0]{\begingroup \@sanitize@url \@href}%
	\providecommand \@href[1]{\@@startlink{#1}\@@href}%
	\providecommand \@@href[1]{\endgroup#1\@@endlink}%
	\providecommand \@sanitize@url [0]{\catcode `\\12\catcode `\$12\catcode
		`\&12\catcode `\#12\catcode `\^12\catcode `\_12\catcode `\%12\relax}%
	\providecommand \@@startlink[1]{}%
	\providecommand \@@endlink[0]{}%
	\providecommand \url  [0]{\begingroup\@sanitize@url \@url }%
	\providecommand \@url [1]{\endgroup\@href {#1}{\urlprefix }}%
	\providecommand \urlprefix  [0]{URL }%
	\providecommand \Eprint [0]{\href }%
	\providecommand \doibase [0]{http://dx.doi.org/}%
	\providecommand \selectlanguage [0]{\@gobble}%
	\providecommand \bibinfo  [0]{\@secondoftwo}%
	\providecommand \bibfield  [0]{\@secondoftwo}%
	\providecommand \translation [1]{[#1]}%
	\providecommand \BibitemOpen [0]{}%
	\providecommand \bibitemStop [0]{}%
	\providecommand \bibitemNoStop [0]{.\EOS\space}%
	\providecommand \EOS [0]{\spacefactor3000\relax}%
	\providecommand \BibitemShut  [1]{\csname bibitem#1\endcsname}%
	\let\auto@bib@innerbib\@empty
	\bibitem [{\citenamefont {Harned}\ and\ \citenamefont
		{Owen}(1943)}]{Harned1943}%
	\BibitemOpen
	\bibfield  {author} {\bibinfo {author} {\bibfnamefont {H.~S.}\ \bibnamefont
			{Harned}}\ and\ \bibinfo {author} {\bibfnamefont {B.~B.}\ \bibnamefont
			{Owen}},\ }\href@noop {} {\emph {\bibinfo {title} {{The Physical Chemistry of
					Electrolyte Solutions}}}}\ (\bibinfo  {publisher} {American Chemical Society
		Monograph Series},\ \bibinfo {year} {1943})\BibitemShut {NoStop}%
	\bibitem [{\citenamefont {Robinson}\ and\ \citenamefont
		{Stokes}(2002)}]{Robinson2002}%
	\BibitemOpen
	\bibfield  {author} {\bibinfo {author} {\bibfnamefont {R.~A.}\ \bibnamefont
			{Robinson}}\ and\ \bibinfo {author} {\bibfnamefont {R.~H.}\ \bibnamefont
			{Stokes}},\ }\href@noop {} {\emph {\bibinfo {title} {{Electrolyte
					Solutions}}}}\ (\bibinfo  {publisher} {Dover Publications},\ \bibinfo {year}
	{2002})\BibitemShut {NoStop}%
	\bibitem [{\citenamefont {Heerema}\ \emph {et~al.}(2015)\citenamefont
		{Heerema}, \citenamefont {Schneider}, \citenamefont {Rozemuller},
		\citenamefont {Vicarelli}, \citenamefont {Zandbergen},\ and\ \citenamefont
		{Dekker}}]{Heerema2015}%
	\BibitemOpen
	\bibfield  {author} {\bibinfo {author} {\bibfnamefont {S.~J.}\ \bibnamefont
			{Heerema}}, \bibinfo {author} {\bibfnamefont {G.~F.}\ \bibnamefont
			{Schneider}}, \bibinfo {author} {\bibfnamefont {M.}~\bibnamefont
			{Rozemuller}}, \bibinfo {author} {\bibfnamefont {L.}~\bibnamefont
			{Vicarelli}}, \bibinfo {author} {\bibfnamefont {H.~W.}\ \bibnamefont
			{Zandbergen}}, \ and\ \bibinfo {author} {\bibfnamefont {C.}~\bibnamefont
			{Dekker}},\ }\href {\doibase 10.1088/0957-4484/26/7/074001} {\bibfield
		{journal} {\bibinfo  {journal} {Nanotechnology}\ }\textbf {\bibinfo {volume}
			{26}},\ \bibinfo {pages} {074001} (\bibinfo {year} {2015})}\BibitemShut
	{NoStop}%
	\bibitem [{\citenamefont {Vasilescu}\ \emph {et~al.}(1974)\citenamefont
		{Vasilescu}, \citenamefont {Teboul}, \citenamefont {Kranck},\ and\
		\citenamefont {Gutmann}}]{VASILESCU1974181}%
	\BibitemOpen
	\bibfield  {author} {\bibinfo {author} {\bibfnamefont {D.}~\bibnamefont
			{Vasilescu}}, \bibinfo {author} {\bibfnamefont {M.}~\bibnamefont {Teboul}},
		\bibinfo {author} {\bibfnamefont {H.}~\bibnamefont {Kranck}}, \ and\ \bibinfo
		{author} {\bibfnamefont {F.}~\bibnamefont {Gutmann}},\ }\href {\doibase
		https://doi.org/10.1016/0013-4686(74)85064-4} {\bibfield  {journal} {\bibinfo
			{journal} {Electrochimica Acta}\ }\textbf {\bibinfo {volume} {19}},\
		\bibinfo {pages} {181} (\bibinfo {year} {1974})}\BibitemShut {NoStop}%
	\bibitem [{\citenamefont {Mathwig}\ \emph {et~al.}(2012)\citenamefont
		{Mathwig}, \citenamefont {Mampallil}, \citenamefont {Kang},\ and\
		\citenamefont {Lemay}}]{Mathwig2012}%
	\BibitemOpen
	\bibfield  {author} {\bibinfo {author} {\bibfnamefont {K.}~\bibnamefont
			{Mathwig}}, \bibinfo {author} {\bibfnamefont {D.}~\bibnamefont {Mampallil}},
		\bibinfo {author} {\bibfnamefont {S.}~\bibnamefont {Kang}}, \ and\ \bibinfo
		{author} {\bibfnamefont {S.~G.}\ \bibnamefont {Lemay}},\ }\href {\doibase
		10.1103/PhysRevLett.109.118302} {\bibfield  {journal} {\bibinfo  {journal}
			{Phys. Rev. Lett.}\ }\textbf {\bibinfo {volume} {109}},\ \bibinfo {pages}
		{118302} (\bibinfo {year} {2012})}\BibitemShut {NoStop}%
	\bibitem [{\citenamefont {Secchi}\ \emph {et~al.}(2016)\citenamefont {Secchi},
		\citenamefont {Niguès}, \citenamefont {Jubin}, \citenamefont {Siria},\ and\
		\citenamefont {Bocquet}}]{Secchi2016}%
	\BibitemOpen
	\bibfield  {author} {\bibinfo {author} {\bibfnamefont {E.}~\bibnamefont
			{Secchi}}, \bibinfo {author} {\bibfnamefont {A.}~\bibnamefont {Niguès}},
		\bibinfo {author} {\bibfnamefont {L.}~\bibnamefont {Jubin}}, \bibinfo
		{author} {\bibfnamefont {A.}~\bibnamefont {Siria}}, \ and\ \bibinfo {author}
		{\bibfnamefont {L.}~\bibnamefont {Bocquet}},\ }\href {\doibase
		10.1103/PhysRevLett.116.154501} {\bibfield  {journal} {\bibinfo  {journal}
			{Phys. Rev. Lett.}\ }\textbf {\bibinfo {volume} {116}},\ \bibinfo {pages}
		{154501} (\bibinfo {year} {2016})}\BibitemShut {NoStop}%
	\bibitem [{\citenamefont {Chandra}\ and\ \citenamefont
		{Bagchi}(2000)}]{Chandra2000a}%
	\BibitemOpen
	\bibfield  {author} {\bibinfo {author} {\bibfnamefont {A.}~\bibnamefont
			{Chandra}}\ and\ \bibinfo {author} {\bibfnamefont {B.}~\bibnamefont
			{Bagchi}},\ }\href {\doibase 10.1063/1.480751} {\bibfield  {journal}
		{\bibinfo  {journal} {J. Chem. Phys.}\ }\textbf {\bibinfo {volume} {112}},\
		\bibinfo {pages} {1876} (\bibinfo {year} {2000})}\BibitemShut {NoStop}%
	\bibitem [{\citenamefont {Yamaguchi}\ \emph {et~al.}(2007)\citenamefont
		{Yamaguchi}, \citenamefont {Matsuoka},\ and\ \citenamefont
		{Koda}}]{Yamaguchi2007}%
	\BibitemOpen
	\bibfield  {author} {\bibinfo {author} {\bibfnamefont {T.}~\bibnamefont
			{Yamaguchi}}, \bibinfo {author} {\bibfnamefont {T.}~\bibnamefont {Matsuoka}},
		\ and\ \bibinfo {author} {\bibfnamefont {S.}~\bibnamefont {Koda}},\ }\href
	{\doibase 10.1063/1.2806289} {\bibfield  {journal} {\bibinfo  {journal} {J.
				Chem. Phys.}\ }\textbf {\bibinfo {volume} {127}},\ \bibinfo {pages} {234501}
		(\bibinfo {year} {2007})}\BibitemShut {NoStop}%
	\bibitem [{\citenamefont {Lu}\ \emph {et~al.}(2015)\citenamefont {Lu},
		\citenamefont {Dean},\ and\ \citenamefont {Podgornik}}]{Dean}%
	\BibitemOpen
	\bibfield  {author} {\bibinfo {author} {\bibfnamefont {B.~S.}\ \bibnamefont
			{Lu}}, \bibinfo {author} {\bibfnamefont {D.~S.}\ \bibnamefont {Dean}}, \ and\
		\bibinfo {author} {\bibfnamefont {R.}~\bibnamefont {Podgornik}},\ }\href
	{\doibase 10.1209/0295-5075/112/20001} {\bibfield  {journal} {\bibinfo
			{journal} {Europhys. Lett.}\ }\textbf {\bibinfo {volume} {112}},\ \bibinfo
		{pages} {20001} (\bibinfo {year} {2015})}\BibitemShut {NoStop}%
	\bibitem [{\citenamefont {Dean}\ and\ \citenamefont
		{Podgornik}(2014)}]{Dean2014}%
	\BibitemOpen
	\bibfield  {author} {\bibinfo {author} {\bibfnamefont {D.~S.}\ \bibnamefont
			{Dean}}\ and\ \bibinfo {author} {\bibfnamefont {R.}~\bibnamefont
			{Podgornik}},\ }\href {\doibase 10.1103/PhysRevE.89.032117} {\bibfield
		{journal} {\bibinfo  {journal} {Phys. Rev. E}\ }\textbf {\bibinfo {volume}
			{89}},\ \bibinfo {pages} {032117} (\bibinfo {year} {2014})}\BibitemShut
	{NoStop}%
	\bibitem [{\citenamefont {P{\'{e}}raud}\ \emph {et~al.}(2017)\citenamefont
		{P{\'{e}}raud}, \citenamefont {Nonaka}, \citenamefont {Bell}, \citenamefont
		{Donev},\ and\ \citenamefont {Garcia}}]{Peraud2017}%
	\BibitemOpen
	\bibfield  {author} {\bibinfo {author} {\bibfnamefont {J.~P.}\ \bibnamefont
			{P{\'{e}}raud}}, \bibinfo {author} {\bibfnamefont {A.~J.}\ \bibnamefont
			{Nonaka}}, \bibinfo {author} {\bibfnamefont {J.~B.}\ \bibnamefont {Bell}},
		\bibinfo {author} {\bibfnamefont {A.}~\bibnamefont {Donev}}, \ and\ \bibinfo
		{author} {\bibfnamefont {A.~L.}\ \bibnamefont {Garcia}},\ }\href {\doibase
		10.1073/pnas.1714464114} {\bibfield  {journal} {\bibinfo  {journal}
			{Proceedings of the National Academy of Sciences of the United States of
				America}\ }\textbf {\bibinfo {volume} {114}},\ \bibinfo {pages} {10829}
		(\bibinfo {year} {2017})}\BibitemShut {NoStop}%
	\bibitem [{\citenamefont {Lee}\ \emph {et~al.}(2018)\citenamefont {Lee},
		\citenamefont {Hansen}, \citenamefont {Bernard},\ and\ \citenamefont
		{Rotenberg}}]{Lee2018}%
	\BibitemOpen
	\bibfield  {author} {\bibinfo {author} {\bibfnamefont {A.~A.}\ \bibnamefont
			{Lee}}, \bibinfo {author} {\bibfnamefont {J.~P.}\ \bibnamefont {Hansen}},
		\bibinfo {author} {\bibfnamefont {O.}~\bibnamefont {Bernard}}, \ and\
		\bibinfo {author} {\bibfnamefont {B.}~\bibnamefont {Rotenberg}},\ }\href
	{\doibase 10.1080/00268976.2018.1478137} {\bibfield  {journal} {\bibinfo
			{journal} {Mol. Phys.}\ }\textbf {\bibinfo {volume} {116}},\ \bibinfo {pages}
		{3147} (\bibinfo {year} {2018})}\BibitemShut {NoStop}%
	\bibitem [{\citenamefont {Mahdisoltani}\ and\ \citenamefont
		{Golestanian}(2021{\natexlab{a}})}]{Mahdisoltani2021a}%
	\BibitemOpen
	\bibfield  {author} {\bibinfo {author} {\bibfnamefont {S.}~\bibnamefont
			{Mahdisoltani}}\ and\ \bibinfo {author} {\bibfnamefont {R.}~\bibnamefont
			{Golestanian}},\ }\href {\doibase 10.1103/PhysRevLett.126.158002} {\bibfield
		{journal} {\bibinfo  {journal} {Phys. Rev. Lett.}\ }\textbf {\bibinfo
			{volume} {126}},\ \bibinfo {pages} {158002} (\bibinfo {year}
		{2021}{\natexlab{a}})}\BibitemShut {NoStop}%
	\bibitem [{\citenamefont {Mahdisoltani}\ and\ \citenamefont
		{Golestanian}(2021{\natexlab{b}})}]{Mahdisoltani2021}%
	\BibitemOpen
	\bibfield  {author} {\bibinfo {author} {\bibfnamefont {S.}~\bibnamefont
			{Mahdisoltani}}\ and\ \bibinfo {author} {\bibfnamefont {R.}~\bibnamefont
			{Golestanian}},\ }\href {\doibase 10.1088/1367-2630/ac0f1a} {\bibfield
		{journal} {\bibinfo  {journal} {New Journal of Physics}\ }\textbf {\bibinfo
			{volume} {23}},\ \bibinfo {pages} {073034} (\bibinfo {year}
		{2021}{\natexlab{b}})}\BibitemShut {NoStop}%
	\bibitem [{\citenamefont {Du}\ \emph {et~al.}(2024)\citenamefont {Du},
		\citenamefont {Dean}, \citenamefont {Miao},\ and\ \citenamefont
		{Podgornik}}]{Du2024}%
	\BibitemOpen
	\bibfield  {author} {\bibinfo {author} {\bibfnamefont {G.}~\bibnamefont
			{Du}}, \bibinfo {author} {\bibfnamefont {D.~S.}\ \bibnamefont {Dean}},
		\bibinfo {author} {\bibfnamefont {B.}~\bibnamefont {Miao}}, \ and\ \bibinfo
		{author} {\bibfnamefont {R.}~\bibnamefont {Podgornik}},\ }\href
	{https://arxiv.org/abs/2404.06028} {} (\bibinfo {year} {2024}),\ \Eprint
	{http://arxiv.org/abs/2404.06028} {arXiv:2404.06028} \BibitemShut {NoStop}%
	\bibitem [{\citenamefont {Kawasaki}(1994)}]{Kawasaki1994}%
	\BibitemOpen
	\bibfield  {author} {\bibinfo {author} {\bibfnamefont {K.}~\bibnamefont
			{Kawasaki}},\ }\href {\doibase 10.1016/0378-4371(94)90533-9} {\bibfield
		{journal} {\bibinfo  {journal} {Physica A}\ }\textbf {\bibinfo {volume}
			{208}},\ \bibinfo {pages} {35} (\bibinfo {year} {1994})}\BibitemShut
	{NoStop}%
	\bibitem [{\citenamefont {Dean}(1996)}]{Dean1996}%
	\BibitemOpen
	\bibfield  {author} {\bibinfo {author} {\bibfnamefont {D.~S.}\ \bibnamefont
			{Dean}},\ }\href {\doibase 10.1088/0305-4470/29/24/001} {\bibfield  {journal}
		{\bibinfo  {journal} {J. Phys. A: Math. Gen.}\ }\textbf {\bibinfo {volume}
			{29}},\ \bibinfo {pages} {L613} (\bibinfo {year} {1996})}\BibitemShut
	{NoStop}%
	\bibitem [{\citenamefont {D{\'{e}}mery}\ and\ \citenamefont
		{Dean}(2015)}]{Demery2015a}%
	\BibitemOpen
	\bibfield  {author} {\bibinfo {author} {\bibfnamefont {V.}~\bibnamefont
			{D{\'{e}}mery}}\ and\ \bibinfo {author} {\bibfnamefont {D.~S.}\ \bibnamefont
			{Dean}},\ }\href {\doibase 10.1088/1742-5468/2016/02/023106} {\bibfield
		{journal} {\bibinfo  {journal} {J. Stat. Mech.}\ }\textbf {\bibinfo {volume}
			{2016}},\ \bibinfo {pages} {023106} (\bibinfo {year} {2015})}\BibitemShut
	{NoStop}%
	\bibitem [{\citenamefont {Péraud}\ \emph {et~al.}(2016)\citenamefont
		{Péraud}, \citenamefont {Nonaka}, \citenamefont {Chaudhri}, \citenamefont
		{Bell}, \citenamefont {Donev},\ and\ \citenamefont {Garcia}}]{Peraud2016}%
	\BibitemOpen
	\bibfield  {author} {\bibinfo {author} {\bibfnamefont {J.-P.}\ \bibnamefont
			{Péraud}}, \bibinfo {author} {\bibfnamefont {A.}~\bibnamefont {Nonaka}},
		\bibinfo {author} {\bibfnamefont {A.}~\bibnamefont {Chaudhri}}, \bibinfo
		{author} {\bibfnamefont {J.~B.}\ \bibnamefont {Bell}}, \bibinfo {author}
		{\bibfnamefont {A.}~\bibnamefont {Donev}}, \ and\ \bibinfo {author}
		{\bibfnamefont {A.~L.}\ \bibnamefont {Garcia}},\ }\href {\doibase
		10.1103/PhysRevFluids.1.074103} {\bibfield  {journal} {\bibinfo  {journal}
			{Physical Review Fluids}\ }\textbf {\bibinfo {volume} {1}},\ \bibinfo {pages}
		{074103} (\bibinfo {year} {2016})}\BibitemShut {NoStop}%
	\bibitem [{\citenamefont {Donev}\ \emph {et~al.}(2019)\citenamefont {Donev},
		\citenamefont {Garcia}, \citenamefont {P{\'{e}}raud}, \citenamefont
		{Nonaka},\ and\ \citenamefont {Bell}}]{Donev2019}%
	\BibitemOpen
	\bibfield  {author} {\bibinfo {author} {\bibfnamefont {A.}~\bibnamefont
			{Donev}}, \bibinfo {author} {\bibfnamefont {A.~L.}\ \bibnamefont {Garcia}},
		\bibinfo {author} {\bibfnamefont {J.~P.}\ \bibnamefont {P{\'{e}}raud}},
		\bibinfo {author} {\bibfnamefont {A.~J.}\ \bibnamefont {Nonaka}}, \ and\
		\bibinfo {author} {\bibfnamefont {J.~B.}\ \bibnamefont {Bell}},\ }\href
	{\doibase 10.1016/j.coelec.2018.09.004} {\bibfield  {journal} {\bibinfo
			{journal} {Current Opinion in Electrochemistry}\ }\textbf {\bibinfo {volume}
			{13}},\ \bibinfo {pages} {1} (\bibinfo {year} {2019})}\BibitemShut {NoStop}%
	\bibitem [{\citenamefont {Avni}\ \emph
		{et~al.}(2022{\natexlab{a}})\citenamefont {Avni}, \citenamefont {Adar},
		\citenamefont {Andelman},\ and\ \citenamefont {Orland}}]{Avni2022}%
	\BibitemOpen
	\bibfield  {author} {\bibinfo {author} {\bibfnamefont {Y.}~\bibnamefont
			{Avni}}, \bibinfo {author} {\bibfnamefont {R.~M.}\ \bibnamefont {Adar}},
		\bibinfo {author} {\bibfnamefont {D.}~\bibnamefont {Andelman}}, \ and\
		\bibinfo {author} {\bibfnamefont {H.}~\bibnamefont {Orland}},\ }\href
	{\doibase 10.1103/PhysRevLett.128.098002} {\bibfield  {journal} {\bibinfo
			{journal} {Phys. Rev. Lett.}\ }\textbf {\bibinfo {volume} {128}},\ \bibinfo
		{pages} {098002} (\bibinfo {year} {2022}{\natexlab{a}})}\BibitemShut
	{NoStop}%
	\bibitem [{\citenamefont {Avni}\ \emph
		{et~al.}(2022{\natexlab{b}})\citenamefont {Avni}, \citenamefont {Andelman},\
		and\ \citenamefont {Orland}}]{Avni2022a}%
	\BibitemOpen
	\bibfield  {author} {\bibinfo {author} {\bibfnamefont {Y.}~\bibnamefont
			{Avni}}, \bibinfo {author} {\bibfnamefont {D.}~\bibnamefont {Andelman}}, \
		and\ \bibinfo {author} {\bibfnamefont {H.}~\bibnamefont {Orland}},\ }\href
	{http://arxiv.org/abs/2207.10116} {\bibfield  {journal} {\bibinfo  {journal}
			{J. Chem. Phys.}\ }\textbf {\bibinfo {volume} {157}},\ \bibinfo {pages}
		{154502} (\bibinfo {year} {2022}{\natexlab{b}})}\BibitemShut {NoStop}%
	\bibitem [{\citenamefont {Bonneau}\ \emph {et~al.}(2023)\citenamefont
		{Bonneau}, \citenamefont {Démery},\ and\ \citenamefont
		{Rapha\"el}}]{Bonneau2023}%
	\BibitemOpen
	\bibfield  {author} {\bibinfo {author} {\bibfnamefont {H.}~\bibnamefont
			{Bonneau}}, \bibinfo {author} {\bibfnamefont {V.}~\bibnamefont {Démery}}, \
		and\ \bibinfo {author} {\bibfnamefont {E.}~\bibnamefont {Rapha\"el}},\ }\href
	{\doibase 10.1088/1742-5468/acdced} {\bibfield  {journal} {\bibinfo
			{journal} {J. Stat. Mech.}\ }\textbf {\bibinfo {volume} {2023}},\ \bibinfo
		{pages} {073205} (\bibinfo {year} {2023})}\BibitemShut {NoStop}%
	\bibitem [{\citenamefont {{Hoang Ngoc}}\ \emph {et~al.}(2023)\citenamefont
		{{Hoang Ngoc}}, \citenamefont {Rotenberg},\ and\ \citenamefont
		{Marbach}}]{Minh2023}%
	\BibitemOpen
	\bibfield  {author} {\bibinfo {author} {\bibfnamefont {M.-T.}\ \bibnamefont
			{{Hoang Ngoc}}}, \bibinfo {author} {\bibfnamefont {B.}~\bibnamefont
			{Rotenberg}}, \ and\ \bibinfo {author} {\bibfnamefont {S.}~\bibnamefont
			{Marbach}},\ }\href {http://arxiv.org/abs/2302.03393} {\bibfield  {journal}
		{\bibinfo  {journal} {Faraday Discussions}\ }\textbf {\bibinfo {volume}
			{246}},\ \bibinfo {pages} {225} (\bibinfo {year} {2023})}\BibitemShut
	{NoStop}%
	\bibitem [{\citenamefont {Berthoumieux}\ \emph {et~al.}(2024)\citenamefont
		{Berthoumieux}, \citenamefont {Démery},\ and\ \citenamefont
		{Maggs}}]{Berthoumieux2024}%
	\BibitemOpen
	\bibfield  {author} {\bibinfo {author} {\bibfnamefont {H.}~\bibnamefont
			{Berthoumieux}}, \bibinfo {author} {\bibfnamefont {V.}~\bibnamefont
			{Démery}}, \ and\ \bibinfo {author} {\bibfnamefont {A.~C.}\ \bibnamefont
			{Maggs}},\ }\href {https://arxiv.org/abs/2405.05882} {} (\bibinfo {year}
	{2024}),\ \Eprint {http://arxiv.org/abs/2405.05882} {arXiv:2405.05882}
	\BibitemShut {NoStop}%
	\bibitem [{\citenamefont {Frusawa}(2020)}]{Frusawa2020}%
	\BibitemOpen
	\bibfield  {author} {\bibinfo {author} {\bibfnamefont {H.}~\bibnamefont
			{Frusawa}},\ }\href {\doibase 10.3390/e22010034} {\bibfield  {journal}
		{\bibinfo  {journal} {Entropy}\ }\textbf {\bibinfo {volume} {22}},\ \bibinfo
		{pages} {34} (\bibinfo {year} {2020})}\BibitemShut {NoStop}%
	\bibitem [{\citenamefont {Frusawa}(2022)}]{Frusawa2022}%
	\BibitemOpen
	\bibfield  {author} {\bibinfo {author} {\bibfnamefont {H.}~\bibnamefont
			{Frusawa}},\ }\href {\doibase 10.1039/d1sm01811f} {\bibfield  {journal}
		{\bibinfo  {journal} {Soft Matter}\ }\textbf {\bibinfo {volume} {18}},\
		\bibinfo {pages} {4280} (\bibinfo {year} {2022})}\BibitemShut {NoStop}%
	\bibitem [{\citenamefont {Wada}(2005)}]{Wada2005}%
	\BibitemOpen
	\bibfield  {author} {\bibinfo {author} {\bibfnamefont {H.}~\bibnamefont
			{Wada}},\ }\href {\doibase 10.1088/1742-5468/2005/01/P01001} {\bibfield
		{journal} {\bibinfo  {journal} {Journal of Statistical Mechanics: Theory and
				Experiment}\ }\textbf {\bibinfo {volume} {2005}},\ \bibinfo {pages} {P01001}
		(\bibinfo {year} {2005})}\BibitemShut {NoStop}%
	\bibitem [{\citenamefont {Okamoto}(2022)}]{Okamoto2022}%
	\BibitemOpen
	\bibfield  {author} {\bibinfo {author} {\bibfnamefont {R.}~\bibnamefont
			{Okamoto}},\ }\href {\doibase 10.1088/1742-5468/ac8c8d} {\bibfield  {journal}
		{\bibinfo  {journal} {Journal of Statistical Mechanics: Theory and
				Experiment}\ }\textbf {\bibinfo {volume} {2022}},\ \bibinfo {pages} {093203}
		(\bibinfo {year} {2022})}\BibitemShut {NoStop}%
	\bibitem [{\citenamefont {Robin}(2024)}]{Robin2024}%
	\BibitemOpen
	\bibfield  {author} {\bibinfo {author} {\bibfnamefont {P.}~\bibnamefont
			{Robin}},\ }\href {\doibase 10.1063/5.0188215} {\bibfield  {journal}
		{\bibinfo  {journal} {J. Chem. Phys.}\ }\textbf {\bibinfo {volume} {160}},\
		\bibinfo {pages} {064503} (\bibinfo {year} {2024})}\BibitemShut {NoStop}%
	\bibitem [{\citenamefont {Déjardin}\ \emph {et~al.}(2018)\citenamefont
		{Déjardin}, \citenamefont {Cornaton}, \citenamefont {Ghesquière},
		\citenamefont {Caliot},\ and\ \citenamefont {Brouzet}}]{Dejardin2018}%
	\BibitemOpen
	\bibfield  {author} {\bibinfo {author} {\bibfnamefont {P.~M.}\ \bibnamefont
			{Déjardin}}, \bibinfo {author} {\bibfnamefont {Y.}~\bibnamefont {Cornaton}},
		\bibinfo {author} {\bibfnamefont {P.}~\bibnamefont {Ghesquière}}, \bibinfo
		{author} {\bibfnamefont {C.}~\bibnamefont {Caliot}}, \ and\ \bibinfo {author}
		{\bibfnamefont {R.}~\bibnamefont {Brouzet}},\ }\href
	{https://pubs.aip.org/jcp/article/148/4/044504/75376/Calculation-of-the-orientational-linear-and}
	{\bibfield  {journal} {\bibinfo  {journal} {J. Chem. Phys.}\ }\textbf
		{\bibinfo {volume} {148}},\ \bibinfo {pages} {044504} (\bibinfo {year}
		{2018})}\BibitemShut {NoStop}%
	\bibitem [{\citenamefont {Bopp}\ \emph {et~al.}(1998)\citenamefont {Bopp},
		\citenamefont {Kornyshev},\ and\ \citenamefont {Sutmann}}]{Bopp1998}%
	\BibitemOpen
	\bibfield  {author} {\bibinfo {author} {\bibfnamefont {P.~A.}\ \bibnamefont
			{Bopp}}, \bibinfo {author} {\bibfnamefont {A.~A.}\ \bibnamefont {Kornyshev}},
		\ and\ \bibinfo {author} {\bibfnamefont {G.}~\bibnamefont {Sutmann}},\ }\href
	{\doibase 10.1063/1.476884} {\bibfield  {journal} {\bibinfo  {journal} {J.
				Chem. Phys.}\ }\textbf {\bibinfo {volume} {109}},\ \bibinfo {pages} {1939}
		(\bibinfo {year} {1998})}\BibitemShut {NoStop}%
	\bibitem [{\citenamefont {Kaatze}\ and\ \citenamefont
		{Uhlendorf}(1981)}]{Kaatze1981}%
	\BibitemOpen
	\bibfield  {author} {\bibinfo {author} {\bibfnamefont {U.}~\bibnamefont
			{Kaatze}}\ and\ \bibinfo {author} {\bibfnamefont {V.}~\bibnamefont
			{Uhlendorf}},\ }\href {\doibase 10.1524/zpch.1981.126.2.151} {\bibfield
		{journal} {\bibinfo  {journal} {Zeitschrift fur Physikalische Chemie}\
		}\textbf {\bibinfo {volume} {126}},\ \bibinfo {pages} {151} (\bibinfo {year}
		{1981})}\BibitemShut {NoStop}%
	\bibitem [{\citenamefont {Afsar}\ and\ \citenamefont
		{Hasted}(1977)}]{Afsar1977}%
	\BibitemOpen
	\bibfield  {author} {\bibinfo {author} {\bibfnamefont {M.~N.}\ \bibnamefont
			{Afsar}}\ and\ \bibinfo {author} {\bibfnamefont {B.}~\bibnamefont {Hasted}},\
	}\href@noop {} {\bibfield  {journal} {\bibinfo  {journal} {J. Opt. Soc. Am.}\
		}\textbf {\bibinfo {volume} {67}},\ \bibinfo {pages} {902} (\bibinfo {year}
		{1977})}\BibitemShut {NoStop}%
	\bibitem [{\citenamefont {Ruse}(1971)}]{Ruse1971}%
	\BibitemOpen
	\bibfield  {author} {\bibinfo {author} {\bibfnamefont {A.~N.}\ \bibnamefont
			{Ruse}},\ }\href@noop {} {\bibfield  {journal} {\bibinfo  {journal} {J. Opt.
				Soc. Am.}\ }\textbf {\bibinfo {volume} {61}},\ \bibinfo {pages} {895}
		(\bibinfo {year} {1971})}\BibitemShut {NoStop}%
	\bibitem [{\citenamefont {Geissler}\ \emph {et~al.}(2001)\citenamefont
		{Geissler}, \citenamefont {Dellago}, \citenamefont {Chandler}, \citenamefont
		{Hutter},\ and\ \citenamefont {Parrinello}}]{Geissler2001}%
	\BibitemOpen
	\bibfield  {author} {\bibinfo {author} {\bibfnamefont {P.~L.}\ \bibnamefont
			{Geissler}}, \bibinfo {author} {\bibfnamefont {C.}~\bibnamefont {Dellago}},
		\bibinfo {author} {\bibfnamefont {D.}~\bibnamefont {Chandler}}, \bibinfo
		{author} {\bibfnamefont {J.}~\bibnamefont {Hutter}}, \ and\ \bibinfo {author}
		{\bibfnamefont {M.}~\bibnamefont {Parrinello}},\ }\href {\doibase
		10.1126/science.1056991} {\bibfield  {journal} {\bibinfo  {journal}
			{Science}\ }\textbf {\bibinfo {volume} {291}},\ \bibinfo {pages} {2121}
		(\bibinfo {year} {2001})}\BibitemShut {NoStop}%
	\bibitem [{\citenamefont {Song}\ \emph {et~al.}(1996)\citenamefont {Song},
		\citenamefont {Chandler},\ and\ \citenamefont {Marcus}}]{Song1996}%
	\BibitemOpen
	\bibfield  {author} {\bibinfo {author} {\bibfnamefont {X.}~\bibnamefont
			{Song}}, \bibinfo {author} {\bibfnamefont {D.}~\bibnamefont {Chandler}}, \
		and\ \bibinfo {author} {\bibfnamefont {R.~A.}\ \bibnamefont {Marcus}},\
	}\href {\doibase 10.1021/jp960887e} {\bibfield  {journal} {\bibinfo
			{journal} {J. Phys. Chem.}\ }\textbf {\bibinfo {volume} {100}},\ \bibinfo
		{pages} {11954} (\bibinfo {year} {1996})}\BibitemShut {NoStop}%
	\bibitem [{\citenamefont {Chubak}\ \emph {et~al.}(2021)\citenamefont {Chubak},
		\citenamefont {Scalfi}, \citenamefont {Carof},\ and\ \citenamefont
		{Rotenberg}}]{Chubak2021}%
	\BibitemOpen
	\bibfield  {author} {\bibinfo {author} {\bibfnamefont {I.}~\bibnamefont
			{Chubak}}, \bibinfo {author} {\bibfnamefont {L.}~\bibnamefont {Scalfi}},
		\bibinfo {author} {\bibfnamefont {A.}~\bibnamefont {Carof}}, \ and\ \bibinfo
		{author} {\bibfnamefont {B.}~\bibnamefont {Rotenberg}},\ }\href {\doibase
		10.1021/acs.jctc.1c00690} {\bibfield  {journal} {\bibinfo  {journal} {J.
				Chem. Th. Comp.}\ }\textbf {\bibinfo {volume} {17}},\ \bibinfo {pages} {6006}
		(\bibinfo {year} {2021})}\BibitemShut {NoStop}%
	\bibitem [{\citenamefont {Chubak}\ \emph {et~al.}(2023)\citenamefont {Chubak},
		\citenamefont {Alon}, \citenamefont {Silletta}, \citenamefont {Madelin},
		\citenamefont {Jerschow},\ and\ \citenamefont {Rotenberg}}]{Chubak2023}%
	\BibitemOpen
	\bibfield  {author} {\bibinfo {author} {\bibfnamefont {I.}~\bibnamefont
			{Chubak}}, \bibinfo {author} {\bibfnamefont {L.}~\bibnamefont {Alon}},
		\bibinfo {author} {\bibfnamefont {E.~V.}\ \bibnamefont {Silletta}}, \bibinfo
		{author} {\bibfnamefont {G.}~\bibnamefont {Madelin}}, \bibinfo {author}
		{\bibfnamefont {A.}~\bibnamefont {Jerschow}}, \ and\ \bibinfo {author}
		{\bibfnamefont {B.}~\bibnamefont {Rotenberg}},\ }\href {\doibase
		10.1038/s41467-022-35695-3} {\bibfield  {journal} {\bibinfo  {journal}
			{Nature Communications}\ }\textbf {\bibinfo {volume} {14}},\ \bibinfo {pages}
		{84} (\bibinfo {year} {2023})}\BibitemShut {NoStop}%
	\bibitem [{\citenamefont {Hubbard}\ and\ \citenamefont
		{Onsager}(1977)}]{Hubbard1977}%
	\BibitemOpen
	\bibfield  {author} {\bibinfo {author} {\bibfnamefont {J.}~\bibnamefont
			{Hubbard}}\ and\ \bibinfo {author} {\bibfnamefont {L.}~\bibnamefont
			{Onsager}},\ }\href {\doibase 10.1063/1.434664} {\bibfield  {journal}
		{\bibinfo  {journal} {J. Chem. Phys.}\ }\textbf {\bibinfo {volume} {67}},\
		\bibinfo {pages} {4850} (\bibinfo {year} {1977})}\BibitemShut {NoStop}%
	\bibitem [{\citenamefont {Hubbard}(1978)}]{Hubbard1978}%
	\BibitemOpen
	\bibfield  {author} {\bibinfo {author} {\bibfnamefont {J.~B.}\ \bibnamefont
			{Hubbard}},\ }\href {\doibase 10.1063/1.435931} {\bibfield  {journal}
		{\bibinfo  {journal} {J. Chem. Phys.}\ }\textbf {\bibinfo {volume} {68}},\
		\bibinfo {pages} {1649} (\bibinfo {year} {1978})}\BibitemShut {NoStop}%
	\bibitem [{\citenamefont {Koneshan}\ \emph {et~al.}(1998)\citenamefont
		{Koneshan}, \citenamefont {Rasaiah}, \citenamefont {Lynden-Bell},\ and\
		\citenamefont {Lee}}]{Koneshan1998}%
	\BibitemOpen
	\bibfield  {author} {\bibinfo {author} {\bibfnamefont {S.}~\bibnamefont
			{Koneshan}}, \bibinfo {author} {\bibfnamefont {J.~C.}\ \bibnamefont
			{Rasaiah}}, \bibinfo {author} {\bibfnamefont {R.~M.}\ \bibnamefont
			{Lynden-Bell}}, \ and\ \bibinfo {author} {\bibfnamefont {S.~H.}\ \bibnamefont
			{Lee}},\ }\href {\doibase 10.1021/jp980642x} {\bibfield  {journal} {\bibinfo
			{journal} {J. Phys. Chem. B}\ }\textbf {\bibinfo {volume} {102}},\ \bibinfo
		{pages} {4193} (\bibinfo {year} {1998})}\BibitemShut {NoStop}%
	\bibitem [{\citenamefont {Banerjee}\ and\ \citenamefont
		{Bagchi}(2019)}]{Banerjee2019a}%
	\BibitemOpen
	\bibfield  {author} {\bibinfo {author} {\bibfnamefont {P.}~\bibnamefont
			{Banerjee}}\ and\ \bibinfo {author} {\bibfnamefont {B.}~\bibnamefont
			{Bagchi}},\ }\href {http://dx.doi.org/10.1063/1.5090765} {\bibfield
		{journal} {\bibinfo  {journal} {J. Chem. Phys.}\ }\textbf {\bibinfo {volume}
			{150}} (\bibinfo {year} {2019})}\BibitemShut {NoStop}%
	\bibitem [{\citenamefont {Sega}\ \emph {et~al.}(2013)\citenamefont {Sega},
		\citenamefont {Kantorovich}, \citenamefont {Arnold},\ and\ \citenamefont
		{Holm}}]{Sega2013}%
	\BibitemOpen
	\bibfield  {author} {\bibinfo {author} {\bibfnamefont {M.}~\bibnamefont
			{Sega}}, \bibinfo {author} {\bibfnamefont {S.~S.}\ \bibnamefont
			{Kantorovich}}, \bibinfo {author} {\bibfnamefont {A.}~\bibnamefont {Arnold}},
		\ and\ \bibinfo {author} {\bibfnamefont {C.}~\bibnamefont {Holm}},\ }in\
	\href {\doibase 10.1007/978-94-007-5012-8_8} {\emph {\bibinfo {booktitle}
			{Recent Advances in Broadband Dielectric Spectroscopy}}},\ \bibinfo {editor}
	{edited by\ \bibinfo {editor} {\bibfnamefont {Y.~P.}\ \bibnamefont
			{Kalmykov}}}\ (\bibinfo  {publisher} {Springer Netherlands},\ \bibinfo {year}
	{2013})\ pp.\ \bibinfo {pages} {103--122}\BibitemShut {NoStop}%
	\bibitem [{\citenamefont {Sega}\ \emph {et~al.}(2014)\citenamefont {Sega},
		\citenamefont {Kantorovich}, \citenamefont {Holm},\ and\ \citenamefont
		{Arnold}}]{Sega2014}%
	\BibitemOpen
	\bibfield  {author} {\bibinfo {author} {\bibfnamefont {M.}~\bibnamefont
			{Sega}}, \bibinfo {author} {\bibfnamefont {S.~S.}\ \bibnamefont
			{Kantorovich}}, \bibinfo {author} {\bibfnamefont {C.}~\bibnamefont {Holm}}, \
		and\ \bibinfo {author} {\bibfnamefont {A.}~\bibnamefont {Arnold}},\ }\href
	{http://dx.doi.org/10.1063/1.4880237} {\bibfield  {journal} {\bibinfo
			{journal} {J. Chem. Phys.}\ }\textbf {\bibinfo {volume} {140}} (\bibinfo
		{year} {2014})}\BibitemShut {NoStop}%
	\bibitem [{\citenamefont {Hoang Ngoc~Minh}\ \emph {et~al.}(2023)\citenamefont
		{Hoang Ngoc~Minh}, \citenamefont {Kim}, \citenamefont {Pireddu},
		\citenamefont {Chubak}, \citenamefont {Nair},\ and\ \citenamefont
		{Rotenberg}}]{Minh2023a}%
	\BibitemOpen
	\bibfield  {author} {\bibinfo {author} {\bibfnamefont {T.}~\bibnamefont
			{Hoang Ngoc~Minh}}, \bibinfo {author} {\bibfnamefont {J.}~\bibnamefont
			{Kim}}, \bibinfo {author} {\bibfnamefont {G.}~\bibnamefont {Pireddu}},
		\bibinfo {author} {\bibfnamefont {I.}~\bibnamefont {Chubak}}, \bibinfo
		{author} {\bibfnamefont {S.}~\bibnamefont {Nair}}, \ and\ \bibinfo {author}
		{\bibfnamefont {B.}~\bibnamefont {Rotenberg}},\ }\href {\doibase
		10.1039/D3FD00026E} {\bibfield  {journal} {\bibinfo  {journal} {Faraday
				Discuss.}\ }\textbf {\bibinfo {volume} {246}},\ \bibinfo {pages} {198}
		(\bibinfo {year} {2023})}\BibitemShut {NoStop}%
	\bibitem [{Note1()}]{Note1}%
	\BibitemOpen
	\bibinfo {note} {Note that we make the choice to use SI units throughout the
		paper: the mapping to cgs or Gaussian units, which are still used by many
		authors, is typically done through the change $\varepsilon _0 \to 1/(4\pi
		)$.}\BibitemShut {Stop}%
	\bibitem [{Note2()}]{Note2}%
	\BibitemOpen
	\bibinfo {note} {See Supplemental Material, which includes Refs. \cite
		{Gardiner1985, Dean1996, Ahmadi2006, Brotto2013, Golestanian2019,
			Berthoumieux2015, Berthoumieux2018, Vatin2021, Maggs2006, Blossey2022,
			Blossey2022a, Becker2023, Hansen1986, Ladanyi1999, Caillol1987, Madden1984,
			Caillol1986, Nakamura2009, Demery2015, Das2013}, for additional information
		about the analytical calculations\label {smlabel}}\BibitemShut {NoStop}%
	\bibitem [{\citenamefont {Doi}\ and\ \citenamefont {Edwards}(1988)}]{Doi1988}%
	\BibitemOpen
	\bibfield  {author} {\bibinfo {author} {\bibfnamefont {M.}~\bibnamefont
			{Doi}}\ and\ \bibinfo {author} {\bibfnamefont {S.~F.}\ \bibnamefont
			{Edwards}},\ }\href@noop {} {\emph {\bibinfo {title} {{The Theory of Polymer
					Dynamics}}}}\ (\bibinfo  {publisher} {Oxford University Press},\ \bibinfo
	{year} {1988})\BibitemShut {NoStop}%
	\bibitem [{\citenamefont {D{\'{e}}mery}\ \emph {et~al.}(2014)\citenamefont
		{D{\'{e}}mery}, \citenamefont {B{\'{e}}nichou},\ and\ \citenamefont
		{Jacquin}}]{Demery2014}%
	\BibitemOpen
	\bibfield  {author} {\bibinfo {author} {\bibfnamefont {V.}~\bibnamefont
			{D{\'{e}}mery}}, \bibinfo {author} {\bibfnamefont {O.}~\bibnamefont
			{B{\'{e}}nichou}}, \ and\ \bibinfo {author} {\bibfnamefont {H.}~\bibnamefont
			{Jacquin}},\ }\href {\doibase 10.1088/1367-2630/16/5/053032} {\bibfield
		{journal} {\bibinfo  {journal} {New J. Phys.}\ }\textbf {\bibinfo {volume}
			{16}},\ \bibinfo {pages} {053032} (\bibinfo {year} {2014})}\BibitemShut
	{NoStop}%
	\bibitem [{\citenamefont {Maggs}\ and\ \citenamefont
		{Everaers}(2006)}]{Maggs2006}%
	\BibitemOpen
	\bibfield  {author} {\bibinfo {author} {\bibfnamefont {A.~C.}\ \bibnamefont
			{Maggs}}\ and\ \bibinfo {author} {\bibfnamefont {R.}~\bibnamefont
			{Everaers}},\ }\href {\doibase 10.1103/PhysRevLett.96.230603} {\bibfield
		{journal} {\bibinfo  {journal} {Phys. Rev. Lett.}\ }\textbf {\bibinfo
			{volume} {96}},\ \bibinfo {pages} {230603} (\bibinfo {year} {2006})},\
	\Eprint {http://arxiv.org/abs/0511623} {0511623} \BibitemShut {NoStop}%
	\bibitem [{\citenamefont {Berthoumieux}\ and\ \citenamefont
		{Maggs}(2015)}]{Berthoumieux2015}%
	\BibitemOpen
	\bibfield  {author} {\bibinfo {author} {\bibfnamefont {H.}~\bibnamefont
			{Berthoumieux}}\ and\ \bibinfo {author} {\bibfnamefont {A.~C.}\ \bibnamefont
			{Maggs}},\ }\href
	{https://pubs.aip.org/jcp/article/143/10/104501/900600/Fluctuation-induced-forces-governed-by-the}
	{\bibfield  {journal} {\bibinfo  {journal} {J. Chem. Phys.}\ }\textbf
		{\bibinfo {volume} {143}},\ \bibinfo {pages} {104501} (\bibinfo {year}
		{2015})}\BibitemShut {NoStop}%
	\bibitem [{\citenamefont {Berthoumieux}(2018)}]{Berthoumieux2018}%
	\BibitemOpen
	\bibfield  {author} {\bibinfo {author} {\bibfnamefont {H.}~\bibnamefont
			{Berthoumieux}},\ }\href
	{https://pubs.aip.org/jcp/article/148/10/104504/197439/Gaussian-field-model-for-polar-fluids-as-a}
	{\bibfield  {journal} {\bibinfo  {journal} {J. Chem. Phys.}\ }\textbf
		{\bibinfo {volume} {148}},\ \bibinfo {pages} {104504} (\bibinfo {year}
		{2018})}\BibitemShut {NoStop}%
	\bibitem [{\citenamefont {Blossey}\ and\ \citenamefont
		{Podgornik}(2022{\natexlab{a}})}]{Blossey2022}%
	\BibitemOpen
	\bibfield  {author} {\bibinfo {author} {\bibfnamefont {R.}~\bibnamefont
			{Blossey}}\ and\ \bibinfo {author} {\bibfnamefont {R.}~\bibnamefont
			{Podgornik}},\ }\href {\doibase 10.1209/0295-5075/ac7d0a} {\bibfield
		{journal} {\bibinfo  {journal} {Europhysics Letters}\ }\textbf {\bibinfo
			{volume} {139}},\ \bibinfo {pages} {27002} (\bibinfo {year}
		{2022}{\natexlab{a}})}\BibitemShut {NoStop}%
	\bibitem [{\citenamefont {Blossey}\ and\ \citenamefont
		{Podgornik}(2022{\natexlab{b}})}]{Blossey2022a}%
	\BibitemOpen
	\bibfield  {author} {\bibinfo {author} {\bibfnamefont {R.}~\bibnamefont
			{Blossey}}\ and\ \bibinfo {author} {\bibfnamefont {R.}~\bibnamefont
			{Podgornik}},\ }\href {\doibase 10.1103/PhysRevResearch.4.023033} {\bibfield
		{journal} {\bibinfo  {journal} {Physical Review Research}\ }\textbf {\bibinfo
			{volume} {4}},\ \bibinfo {pages} {023033} (\bibinfo {year}
		{2022}{\natexlab{b}})}\BibitemShut {NoStop}%
	\bibitem [{\citenamefont {Becker}\ \emph {et~al.}(2023)\citenamefont {Becker},
		\citenamefont {Loche}, \citenamefont {Bonthuis}, \citenamefont {Mouhanna},
		\citenamefont {Netz},\ and\ \citenamefont {Berthoumieux}}]{Becker2023}%
	\BibitemOpen
	\bibfield  {author} {\bibinfo {author} {\bibfnamefont {M.~R.}\ \bibnamefont
			{Becker}}, \bibinfo {author} {\bibfnamefont {P.}~\bibnamefont {Loche}},
		\bibinfo {author} {\bibfnamefont {D.~J.}\ \bibnamefont {Bonthuis}}, \bibinfo
		{author} {\bibfnamefont {D.}~\bibnamefont {Mouhanna}}, \bibinfo {author}
		{\bibfnamefont {R.~R.}\ \bibnamefont {Netz}}, \ and\ \bibinfo {author}
		{\bibfnamefont {H.}~\bibnamefont {Berthoumieux}},\ }\href
	{https://arxiv.org/abs/2303.14846} {} (\bibinfo {year} {2023}),\ \Eprint
	{http://arxiv.org/abs/2303.14846} {arXiv:2303.14846} \BibitemShut {NoStop}%
	\bibitem [{Note3()}]{Note3}%
	\BibitemOpen
	\bibinfo {note} {Throughout the paper, we will use the following convention
		for spatial Fourier transforms: $\protect \tilde {f}(\protect \boldsymbol
		{q}) = \DOTSI \intop \ilimits@ \protect \mathrm {d}\protect \boldsymbol
		{r}\protect \tmspace +\thickmuskip {.2777em} \protect \mathrm {e}^{-\protect
			\mathrm {i}\protect \boldsymbol {q}\cdot \protect \boldsymbol {r}} f(\protect
		\boldsymbol {r})$, and $f(\protect \boldsymbol {r},t) = \protect \frac
		{1}{(2\pi )^3}\DOTSI \intop \ilimits@ \protect \mathrm {d}\protect
		\boldsymbol {q}\protect \tmspace +\thickmuskip {.2777em} \protect \mathrm
		{e}^{\protect \mathrm {i}\protect \boldsymbol {q}\cdot \protect \boldsymbol
			{r}} \protect \tilde {f}(\protect \boldsymbol {q})$}\BibitemShut {NoStop}%
	\bibitem [{\citenamefont {Hansen}\ and\ \citenamefont
		{McDonald}(1986)}]{Hansen1986}%
	\BibitemOpen
	\bibfield  {author} {\bibinfo {author} {\bibfnamefont {J.~P.}\ \bibnamefont
			{Hansen}}\ and\ \bibinfo {author} {\bibfnamefont {I.~R.}\ \bibnamefont
			{McDonald}},\ }\href@noop {} {\emph {\bibinfo {title} {{Theory of simple
					liquids}}}},\ \bibinfo {edition} {2nd}\ ed.\ (\bibinfo  {publisher} {Academic
		Press},\ \bibinfo {year} {1986})\BibitemShut {NoStop}%
	\bibitem [{\citenamefont {Martin}\ and\ \citenamefont
		{Gruber}(1983)}]{Martin1983}%
	\BibitemOpen
	\bibfield  {author} {\bibinfo {author} {\bibfnamefont {P.~A.}\ \bibnamefont
			{Martin}}\ and\ \bibinfo {author} {\bibfnamefont {C.}~\bibnamefont
			{Gruber}},\ }\href {\doibase 10.1007/BF01019506} {\bibfield  {journal}
		{\bibinfo  {journal} {Journal of Statistical Physics}\ }\textbf {\bibinfo
			{volume} {31}},\ \bibinfo {pages} {691} (\bibinfo {year} {1983})}\BibitemShut
	{NoStop}%
	\bibitem [{\citenamefont {Caillol}\ \emph {et~al.}(1986)\citenamefont
		{Caillol}, \citenamefont {Levesque},\ and\ \citenamefont
		{Weis}}]{Caillol1986}%
	\BibitemOpen
	\bibfield  {author} {\bibinfo {author} {\bibfnamefont {J.~M.}\ \bibnamefont
			{Caillol}}, \bibinfo {author} {\bibfnamefont {D.}~\bibnamefont {Levesque}}, \
		and\ \bibinfo {author} {\bibfnamefont {J.~J.}\ \bibnamefont {Weis}},\ }\href
	{\doibase 10.1063/1.451446} {\bibfield  {journal} {\bibinfo  {journal} {J.
				Chem. Phys.}\ }\textbf {\bibinfo {volume} {85}},\ \bibinfo {pages} {6645}
		(\bibinfo {year} {1986})}\BibitemShut {NoStop}%
	\bibitem [{\citenamefont {Madden}\ and\ \citenamefont
		{Kivelson}(1984)}]{Madden1984}%
	\BibitemOpen
	\bibfield  {author} {\bibinfo {author} {\bibfnamefont {P.}~\bibnamefont
			{Madden}}\ and\ \bibinfo {author} {\bibfnamefont {D.}~\bibnamefont
			{Kivelson}},\ }\href {\doibase 10.1002/9780470142806.ch5} {\bibfield
		{journal} {\bibinfo  {journal} {Adv. Chem. Phys.}\ }\textbf {\bibinfo
			{volume} {56}},\ \bibinfo {pages} {467} (\bibinfo {year} {1984})}\BibitemShut
	{NoStop}%
	\bibitem [{\citenamefont {Ladanyi}\ and\ \citenamefont
		{Perng}(1999)}]{Ladanyi1999}%
	\BibitemOpen
	\bibfield  {author} {\bibinfo {author} {\bibfnamefont {B.~M.}\ \bibnamefont
			{Ladanyi}}\ and\ \bibinfo {author} {\bibfnamefont {B.-C.}\ \bibnamefont
			{Perng}},\ }\href {\doibase 10.1063/1.1301531} {\bibfield  {journal}
		{\bibinfo  {journal} {AIP Conference Proceedings}\ }\textbf {\bibinfo
			{volume} {492}},\ \bibinfo {pages} {250} (\bibinfo {year}
		{1999})}\BibitemShut {NoStop}%
	\bibitem [{\citenamefont {Caillol}(1987)}]{Caillol1987}%
	\BibitemOpen
	\bibfield  {author} {\bibinfo {author} {\bibfnamefont {J.~M.}\ \bibnamefont
			{Caillol}},\ }\href {\doibase 10.1209/0295-5075/4/2/006} {\bibfield
		{journal} {\bibinfo  {journal} {Europhys. Lett.}\ }\textbf {\bibinfo {volume}
			{4}},\ \bibinfo {pages} {159} (\bibinfo {year} {1987})}\BibitemShut {NoStop}%
	\bibitem [{\citenamefont {Levesque}\ \emph {et~al.}(1990)\citenamefont
		{Levesque}, \citenamefont {Caillol},\ and\ \citenamefont
		{Weis}}]{Levesque1990}%
	\BibitemOpen
	\bibfield  {author} {\bibinfo {author} {\bibfnamefont {D.}~\bibnamefont
			{Levesque}}, \bibinfo {author} {\bibfnamefont {J.~M.}\ \bibnamefont
			{Caillol}}, \ and\ \bibinfo {author} {\bibfnamefont {J.~J.}\ \bibnamefont
			{Weis}},\ }\href {\doibase 10.1088/0953-8984/2/S/018} {\bibfield  {journal}
		{\bibinfo  {journal} {J. Phys.: Cond. Matt.}\ }\textbf {\bibinfo {volume}
			{2}},\ \bibinfo {pages} {143} (\bibinfo {year} {1990})}\BibitemShut {NoStop}%
	\bibitem [{\citenamefont {Cox}\ and\ \citenamefont {Sprik}(2019)}]{Cox2019}%
	\BibitemOpen
	\bibfield  {author} {\bibinfo {author} {\bibfnamefont {S.~J.}\ \bibnamefont
			{Cox}}\ and\ \bibinfo {author} {\bibfnamefont {M.}~\bibnamefont {Sprik}},\
	}\href {\doibase 10.1063/1.5099207} {\bibfield  {journal} {\bibinfo
			{journal} {J. Chem. Phys.}\ }\textbf {\bibinfo {volume} {151}},\ \bibinfo
		{pages} {064506} (\bibinfo {year} {2019})}\BibitemShut {NoStop}%
	\bibitem [{Note4()}]{Note4}%
	\BibitemOpen
	\bibinfo {note} {The longitudinal permittivity is the one measured in
		experiments. Note that the transverse permittivity can be computed through
		the relation $\varepsilon _T(\omega ) = 1+\chi _T(\omega )$, and has a
		characteristic time $\tau _S/\varepsilon _\protect \text {w}$, as expected
		\cite {Gekle2012}}\BibitemShut {NoStop}%
	\bibitem [{\citenamefont {Abrashkin}\ \emph {et~al.}(2007)\citenamefont
		{Abrashkin}, \citenamefont {Andelman},\ and\ \citenamefont
		{Orland}}]{Abrashkin2007}%
	\BibitemOpen
	\bibfield  {author} {\bibinfo {author} {\bibfnamefont {A.}~\bibnamefont
			{Abrashkin}}, \bibinfo {author} {\bibfnamefont {D.}~\bibnamefont {Andelman}},
		\ and\ \bibinfo {author} {\bibfnamefont {H.}~\bibnamefont {Orland}},\
	}\href@noop {} {\bibfield  {journal} {\bibinfo  {journal} {Phys. Rev. Lett.}\
		}\textbf {\bibinfo {volume} {99}},\ \bibinfo {pages} {077801} (\bibinfo
		{year} {2007})}\BibitemShut {NoStop}%
	\bibitem [{\citenamefont {Silvestrelli}\ and\ \citenamefont
		{Parrinello}(1999)}]{Silvestrelli1999}%
	\BibitemOpen
	\bibfield  {author} {\bibinfo {author} {\bibfnamefont {P.~L.}\ \bibnamefont
			{Silvestrelli}}\ and\ \bibinfo {author} {\bibfnamefont {M.}~\bibnamefont
			{Parrinello}},\ }\href {\doibase 10.1103/PhysRevLett.82.3308} {\bibfield
		{journal} {\bibinfo  {journal} {Phys. Rev. Lett.}\ }\textbf {\bibinfo
			{volume} {82}},\ \bibinfo {pages} {3308} (\bibinfo {year}
		{1999})}\BibitemShut {NoStop}%
	\bibitem [{\citenamefont {Hynes}\ and\ \citenamefont
		{Wolynes}(1981)}]{Hynes1981}%
	\BibitemOpen
	\bibfield  {author} {\bibinfo {author} {\bibfnamefont {J.~T.}\ \bibnamefont
			{Hynes}}\ and\ \bibinfo {author} {\bibfnamefont {P.~G.}\ \bibnamefont
			{Wolynes}},\ }\href {\doibase 10.1063/1.441796} {\bibfield  {journal}
		{\bibinfo  {journal} {The Journal of Chemical Physics}\ }\textbf {\bibinfo
			{volume} {75}},\ \bibinfo {pages} {395} (\bibinfo {year} {1981})}\BibitemShut
	{NoStop}%
	\bibitem [{\citenamefont {Perng}\ and\ \citenamefont
		{Ladanyi}(1998)}]{Perng1998}%
	\BibitemOpen
	\bibfield  {author} {\bibinfo {author} {\bibfnamefont {B.-C.}\ \bibnamefont
			{Perng}}\ and\ \bibinfo {author} {\bibfnamefont {B.~M.}\ \bibnamefont
			{Ladanyi}},\ }\href {\doibase 10.1063/1.476606} {\bibfield  {journal}
		{\bibinfo  {journal} {The Journal of Chemical Physics}\ }\textbf {\bibinfo
			{volume} {109}},\ \bibinfo {pages} {676} (\bibinfo {year}
		{1998})}\BibitemShut {NoStop}%
	\bibitem [{\citenamefont {Maroncelli}\ and\ \citenamefont
		{Fleming}(1988)}]{Maroncelli1988}%
	\BibitemOpen
	\bibfield  {author} {\bibinfo {author} {\bibfnamefont {M.}~\bibnamefont
			{Maroncelli}}\ and\ \bibinfo {author} {\bibfnamefont {G.~R.}\ \bibnamefont
			{Fleming}},\ }\href {\doibase 10.1063/1.455649} {\bibfield  {journal}
		{\bibinfo  {journal} {The Journal of Chemical Physics}\ }\textbf {\bibinfo
			{volume} {89}},\ \bibinfo {pages} {5044} (\bibinfo {year}
		{1988})}\BibitemShut {NoStop}%
	\bibitem [{\citenamefont {Jimenez}\ \emph {et~al.}(1994)\citenamefont
		{Jimenez}, \citenamefont {Fleming}, \citenamefont {Kumar},\ and\
		\citenamefont {Maroncelli}}]{Jimenez1994}%
	\BibitemOpen
	\bibfield  {author} {\bibinfo {author} {\bibfnamefont {R.}~\bibnamefont
			{Jimenez}}, \bibinfo {author} {\bibfnamefont {G.~R.}\ \bibnamefont
			{Fleming}}, \bibinfo {author} {\bibfnamefont {P.~V.}\ \bibnamefont {Kumar}},
		\ and\ \bibinfo {author} {\bibfnamefont {M.}~\bibnamefont {Maroncelli}},\
	}\href {\doibase 10.1038/369471a0} {\bibfield  {journal} {\bibinfo  {journal}
			{Nature}\ }\textbf {\bibinfo {volume} {369}},\ \bibinfo {pages} {471}
		(\bibinfo {year} {1994})}\BibitemShut {NoStop}%
	\bibitem [{\citenamefont {Geissler}\ and\ \citenamefont
		{Chandler}(2000)}]{Geissler2000}%
	\BibitemOpen
	\bibfield  {author} {\bibinfo {author} {\bibfnamefont {P.~L.}\ \bibnamefont
			{Geissler}}\ and\ \bibinfo {author} {\bibfnamefont {D.}~\bibnamefont
			{Chandler}},\ }\href {\doibase 10.1063/1.1290136} {\bibfield  {journal}
		{\bibinfo  {journal} {The Journal of Chemical Physics}\ }\textbf {\bibinfo
			{volume} {113}},\ \bibinfo {pages} {9759} (\bibinfo {year}
		{2000})}\BibitemShut {NoStop}%
	\bibitem [{\citenamefont {Roy}\ \emph {et~al.}(2015)\citenamefont {Roy},
		\citenamefont {Yashonath},\ and\ \citenamefont {Bagchi}}]{Roy2015}%
	\BibitemOpen
	\bibfield  {author} {\bibinfo {author} {\bibfnamefont {S.}~\bibnamefont
			{Roy}}, \bibinfo {author} {\bibfnamefont {S.}~\bibnamefont {Yashonath}}, \
		and\ \bibinfo {author} {\bibfnamefont {B.}~\bibnamefont {Bagchi}},\ }\href
	{\doibase 10.1063/1.4915274} {\bibfield  {journal} {\bibinfo  {journal} {J.
				Chem. Phys.}\ }\textbf {\bibinfo {volume} {142}} (\bibinfo {year} {2015}),\
		10.1063/1.4915274}\BibitemShut {NoStop}%
	\bibitem [{\citenamefont {Dufr{\^{e}}che}\ \emph {et~al.}(2005)\citenamefont
		{Dufr{\^{e}}che}, \citenamefont {Bernard}, \citenamefont {Durand-Vidal},\
		and\ \citenamefont {Turq}}]{Dufreche2005}%
	\BibitemOpen
	\bibfield  {author} {\bibinfo {author} {\bibfnamefont {J.~F.}\ \bibnamefont
			{Dufr{\^{e}}che}}, \bibinfo {author} {\bibfnamefont {O.}~\bibnamefont
			{Bernard}}, \bibinfo {author} {\bibfnamefont {S.}~\bibnamefont
			{Durand-Vidal}}, \ and\ \bibinfo {author} {\bibfnamefont {P.}~\bibnamefont
			{Turq}},\ }\href {\doibase 10.1021/jp050387y} {\bibfield  {journal} {\bibinfo
			{journal} {Journal of Physical Chemistry B}\ }\textbf {\bibinfo {volume}
			{109}},\ \bibinfo {pages} {9873} (\bibinfo {year} {2005})}\BibitemShut
	{NoStop}%
	\bibitem [{\citenamefont {Bernard}\ \emph {et~al.}(2023)\citenamefont
		{Bernard}, \citenamefont {Jardat}, \citenamefont {Rotenberg},\ and\
		\citenamefont {Illien}}]{Bernard2023}%
	\BibitemOpen
	\bibfield  {author} {\bibinfo {author} {\bibfnamefont {O.}~\bibnamefont
			{Bernard}}, \bibinfo {author} {\bibfnamefont {M.}~\bibnamefont {Jardat}},
		\bibinfo {author} {\bibfnamefont {B.}~\bibnamefont {Rotenberg}}, \ and\
		\bibinfo {author} {\bibfnamefont {P.}~\bibnamefont {Illien}},\ }\href
	{\doibase 10.1063/5.0165533} {\bibfield  {journal} {\bibinfo  {journal} {J.
				Chem. Phys.}\ }\textbf {\bibinfo {volume} {159}},\ \bibinfo {pages} {164105}
		(\bibinfo {year} {2023})}\BibitemShut {NoStop}%
	\bibitem [{\citenamefont {Bopp}\ \emph {et~al.}(1996)\citenamefont {Bopp},
		\citenamefont {Kornyshev},\ and\ \citenamefont {Sutmann}}]{Bopp1996}%
	\BibitemOpen
	\bibfield  {author} {\bibinfo {author} {\bibfnamefont {P.~A.}\ \bibnamefont
			{Bopp}}, \bibinfo {author} {\bibfnamefont {A.~A.}\ \bibnamefont {Kornyshev}},
		\ and\ \bibinfo {author} {\bibfnamefont {G.}~\bibnamefont {Sutmann}},\ }\href
	{\doibase 10.1103/PhysRevLett.76.1280} {\bibfield  {journal} {\bibinfo
			{journal} {Phys. Rev. Lett.}\ }\textbf {\bibinfo {volume} {76}},\ \bibinfo
		{pages} {1280} (\bibinfo {year} {1996})}\BibitemShut {NoStop}%
	\bibitem [{\citenamefont {Spinney}\ and\ \citenamefont
		{Morris}(2024)}]{Spinney2024}%
	\BibitemOpen
	\bibfield  {author} {\bibinfo {author} {\bibfnamefont {R.~E.}\ \bibnamefont
			{Spinney}}\ and\ \bibinfo {author} {\bibfnamefont {R.~G.}\ \bibnamefont
			{Morris}},\ }\href {http://arxiv.org/abs/2404.02487} {\  (\bibinfo {year}
		{2024})},\ \bibinfo {note} {arXiv:2404.02487 [cond-mat]}\BibitemShut
	{NoStop}%
	\bibitem [{\citenamefont {Levy}\ \emph {et~al.}(2012)\citenamefont {Levy},
		\citenamefont {Andelman},\ and\ \citenamefont {Orland}}]{Levy2012}%
	\BibitemOpen
	\bibfield  {author} {\bibinfo {author} {\bibfnamefont {A.}~\bibnamefont
			{Levy}}, \bibinfo {author} {\bibfnamefont {D.}~\bibnamefont {Andelman}}, \
		and\ \bibinfo {author} {\bibfnamefont {H.}~\bibnamefont {Orland}},\ }\href
	{\doibase 10.1103/PhysRevLett.108.227801} {\bibfield  {journal} {\bibinfo
			{journal} {Physical Review Letters}\ }\textbf {\bibinfo {volume} {108}},\
		\bibinfo {pages} {227801} (\bibinfo {year} {2012})}\BibitemShut {NoStop}%
	\bibitem [{\citenamefont {Adar}\ \emph {et~al.}(2018)\citenamefont {Adar},
		\citenamefont {Markovich}, \citenamefont {Levy}, \citenamefont {Orland},\
		and\ \citenamefont {Andelman}}]{Adar2018}%
	\BibitemOpen
	\bibfield  {author} {\bibinfo {author} {\bibfnamefont {R.~M.}\ \bibnamefont
			{Adar}}, \bibinfo {author} {\bibfnamefont {T.}~\bibnamefont {Markovich}},
		\bibinfo {author} {\bibfnamefont {A.}~\bibnamefont {Levy}}, \bibinfo {author}
		{\bibfnamefont {H.}~\bibnamefont {Orland}}, \ and\ \bibinfo {author}
		{\bibfnamefont {D.}~\bibnamefont {Andelman}},\ }\href {\doibase
		10.1063/1.5042235} {\bibfield  {journal} {\bibinfo  {journal} {J. Chem.
				Phys.}\ }\textbf {\bibinfo {volume} {149}},\ \bibinfo {pages} {054504}
		(\bibinfo {year} {2018})}\BibitemShut {NoStop}%
	\bibitem [{\citenamefont {Gardiner}(1985)}]{Gardiner1985}%
	\BibitemOpen
	\bibfield  {author} {\bibinfo {author} {\bibfnamefont {C.~W.}\ \bibnamefont
			{Gardiner}},\ }\href@noop {} {\emph {\bibinfo {title} {{Handbook of
					Stochastic Methods}}}}\ (\bibinfo  {publisher} {Springer},\ \bibinfo {year}
	{1985})\BibitemShut {NoStop}%
	\bibitem [{\citenamefont {Ahmadi}\ \emph {et~al.}(2006)\citenamefont {Ahmadi},
		\citenamefont {Marchetti},\ and\ \citenamefont {Liverpool}}]{Ahmadi2006}%
	\BibitemOpen
	\bibfield  {author} {\bibinfo {author} {\bibfnamefont {A.}~\bibnamefont
			{Ahmadi}}, \bibinfo {author} {\bibfnamefont {M.~C.}\ \bibnamefont
			{Marchetti}}, \ and\ \bibinfo {author} {\bibfnamefont {T.~B.}\ \bibnamefont
			{Liverpool}},\ }\href {\doibase 10.1103/PhysRevE.74.061913} {\bibfield
		{journal} {\bibinfo  {journal} {Phys. Rev. E}\ }\textbf {\bibinfo {volume}
			{74}},\ \bibinfo {pages} {061913} (\bibinfo {year} {2006})}\BibitemShut
	{NoStop}%
	\bibitem [{\citenamefont {Brotto}\ \emph {et~al.}(2013)\citenamefont {Brotto},
		\citenamefont {Caussin}, \citenamefont {Lauga},\ and\ \citenamefont
		{Bartolo}}]{Brotto2013}%
	\BibitemOpen
	\bibfield  {author} {\bibinfo {author} {\bibfnamefont {T.}~\bibnamefont
			{Brotto}}, \bibinfo {author} {\bibfnamefont {J.~B.}\ \bibnamefont {Caussin}},
		\bibinfo {author} {\bibfnamefont {E.}~\bibnamefont {Lauga}}, \ and\ \bibinfo
		{author} {\bibfnamefont {D.}~\bibnamefont {Bartolo}},\ }\href {\doibase
		10.1103/PhysRevLett.110.038101} {\bibfield  {journal} {\bibinfo  {journal}
			{Phys. Rev. Lett.}\ }\textbf {\bibinfo {volume} {110}},\ \bibinfo {pages}
		{038101} (\bibinfo {year} {2013})}\BibitemShut {NoStop}%
	\bibitem [{\citenamefont {Golestanian}(2019)}]{Golestanian2019}%
	\BibitemOpen
	\bibfield  {author} {\bibinfo {author} {\bibfnamefont {R.}~\bibnamefont
			{Golestanian}},\ }\href {http://arxiv.org/abs/1909.03747} {\bibfield
		{journal} {\bibinfo  {journal} {Lecture notes from Les Houches Summer School
				'Active Matter and Non-equilibrium Statistical Physics'}\ } (\bibinfo {year}
		{2019})},\ \Eprint {http://arxiv.org/abs/1909.03747} {arXiv:1909.03747}
	\BibitemShut {NoStop}%
	\bibitem [{\citenamefont {Vatin}\ \emph {et~al.}(2021)\citenamefont {Vatin},
		\citenamefont {Porro}, \citenamefont {Sator}, \citenamefont {Dufrêche},\
		and\ \citenamefont {Berthoumieux}}]{Vatin2021}%
	\BibitemOpen
	\bibfield  {author} {\bibinfo {author} {\bibfnamefont {M.}~\bibnamefont
			{Vatin}}, \bibinfo {author} {\bibfnamefont {A.}~\bibnamefont {Porro}},
		\bibinfo {author} {\bibfnamefont {N.}~\bibnamefont {Sator}}, \bibinfo
		{author} {\bibfnamefont {J.~F.}\ \bibnamefont {Dufrêche}}, \ and\ \bibinfo
		{author} {\bibfnamefont {H.}~\bibnamefont {Berthoumieux}},\ }\href
	{https://doi.org/10.1080/00268976.2020.1825849} {\bibfield  {journal}
		{\bibinfo  {journal} {Molecular Physics}\ }\textbf {\bibinfo {volume} {119}}
		(\bibinfo {year} {2021})}\BibitemShut {NoStop}%
	\bibitem [{\citenamefont {Nakamura}\ and\ \citenamefont
		{Yoshimori}(2009)}]{Nakamura2009}%
	\BibitemOpen
	\bibfield  {author} {\bibinfo {author} {\bibfnamefont {T.}~\bibnamefont
			{Nakamura}}\ and\ \bibinfo {author} {\bibfnamefont {A.}~\bibnamefont
			{Yoshimori}},\ }\href {\doibase 10.1088/1751-8113/42/6/065001} {\bibfield
		{journal} {\bibinfo  {journal} {J. of Phys. A}\ }\textbf {\bibinfo {volume}
			{42}},\ \bibinfo {pages} {065001} (\bibinfo {year} {2009})}\BibitemShut
	{NoStop}%
	\bibitem [{\citenamefont {D{\'{e}}mery}(2015)}]{Demery2015}%
	\BibitemOpen
	\bibfield  {author} {\bibinfo {author} {\bibfnamefont {V.}~\bibnamefont
			{D{\'{e}}mery}},\ }\href {\doibase 10.1103/PhysRevE.91.062301} {\bibfield
		{journal} {\bibinfo  {journal} {Phys. Rev. E}\ }\textbf {\bibinfo {volume}
			{91}},\ \bibinfo {pages} {062301} (\bibinfo {year} {2015})}\BibitemShut
	{NoStop}%
	\bibitem [{\citenamefont {Das}\ and\ \citenamefont
		{Yoshimori}(2013)}]{Das2013}%
	\BibitemOpen
	\bibfield  {author} {\bibinfo {author} {\bibfnamefont {S.~P.}\ \bibnamefont
			{Das}}\ and\ \bibinfo {author} {\bibfnamefont {A.}~\bibnamefont
			{Yoshimori}},\ }\href {\doibase 10.1103/PhysRevE.88.043008} {\bibfield
		{journal} {\bibinfo  {journal} {Phys. Rev. E}\ }\textbf {\bibinfo {volume}
			{88}},\ \bibinfo {pages} {043008} (\bibinfo {year} {2013})}\BibitemShut
	{NoStop}%
	\bibitem [{\citenamefont {Gekle}\ and\ \citenamefont {Netz}(2012)}]{Gekle2012}%
	\BibitemOpen
	\bibfield  {author} {\bibinfo {author} {\bibfnamefont {S.}~\bibnamefont
			{Gekle}}\ and\ \bibinfo {author} {\bibfnamefont {R.~R.}\ \bibnamefont
			{Netz}},\ }\href
	{https://pubs.aip.org/jcp/article/137/10/104704/191937/Anisotropy-in-the-dielectric-spectrum-of-hydration}
	{\bibfield  {journal} {\bibinfo  {journal} {J. Chem. Phys.}\ }\textbf
		{\bibinfo {volume} {137}},\ \bibinfo {pages} {104704} (\bibinfo {year}
		{2012})}\BibitemShut {NoStop}%
\end{thebibliography}

%

\end{document}


\onecolumngrid

\beginsupplement

\begin{center}

{\bf \large Stochastic density functional theory for ions in a polar solvent}

$ \ $

Pierre Illien,$^1$ Antoine Carof,$^2$  and Benjamin Rotenberg$^{1,3}$

\textit{$^1$Sorbonne Universit\'e, CNRS, Physico-Chimie des \'Electrolytes et\\ Nanosyst\`emes Interfaciaux (PHENIX), 4 Place Jussieu, 75005 Paris, France}

\textit{$^2$Université de Lorraine, CNRS, Laboratoire de Physique et Chimie Théoriques, Nancy, France}

\textit{$^3$R\'eseau sur le Stockage Electrochimique de l'Energie (RS2E), FR CNRS 3459, 80039 Amiens Cedex, France}

$ \ $

\textbf{\textit{\large Supplemental Material}}
\end{center}

\tableofcontents

\section{Details on stochastic density field theory}

\subsection{Dean-Kawasaki equation for the  density of solvent molecules $n_S(\rr,\uu,t)$}

The starting point of this derivation is the set of overdamped Langevin equations satisfied by $\rr_j^S$ and $\uu_j$ [Eqs.~\eqref{langevin_rs} and~\eqref{langevin_ua} in the main text]. In this Section, we derive the evolution equation of the density of the solvent. We recall its definition:
\begin{equation}
	n_S(\rr,\uu,t) = \sum_{j=1}^{N_S} \underbrace{ \delta(\rr-\rr_j^S(t))\delta(\uu-\uu_j(t))}_{\equiv n_j(\rr,\uu,t)}.
\end{equation}
In this equation, we also define the density associated to one solvent molecule $n_j(\rr,\uu,t)$. We introduce a test function $f(\rr,\uu)$ and use the definition of $n_j$ to write:
\begin{equation}
	\int \dd \rr \int \dd \uu \; n_j(\rr,\uu,t) f(\rr,\uu) = f(\rr_j,\uu_j).
\end{equation}
Deriving with respect to $t$ yields:
\begin{equation}
	\int \dd \rr \int \dd \uu \; \partial_t n_j(\rr,\uu,t)f(\rr,\uu) = \frac{\dd }{\dd t}f(\rr_j,\uu_j).
	\label{testfun}
\end{equation}
Using an extension of It\^o's lemma to function of several variables~\cite{Gardiner1985}, one of them being a unit vector (see Section~\ref{app_Ito}), we get:
\begin{align}
	\frac{\dd }{\dd t}f(\rr_j,\uu_j) =& \sqrt{2D_S } \boldsymbol{\xi}_j^S\cdot \nabla_{\rr_j} f + \sqrt{2D_S^r} \boldsymbol{\zeta}_j^S\cdot \rotop_{\uu_j} f + D_S\nabla^2_{\rr_j}f + D_S^r \rotop_{\uu_j}^2 f \nonumber\\
	&+[\mu_S p (\uu_j\cdot \nabla \EE(\rr_j)]\cdot \nabla_{\rr_j} f + [\mu_S^r p \uu_j \times \EE(\rr_j)]\cdot \rotop_{\uu_j} f \\
	=& \int \dd\rr \int\dd \uu \; \underbrace{\delta(\rr-\rr_j^S(t))\delta(\uu-\uu_j(t))}_{\equiv n_j(\rr,\uu,t)} \Big\{ \sqrt{2D_S } \boldsymbol{\xi}_j^S\cdot \nabla_{\rr} f + \sqrt{2D_S^r} \boldsymbol{\zeta}_j^S\cdot \rotop_{\uu}f + D_S\nabla^2_{\rr}f + D_S^r \rotop_{\uu}^2 f   \nonumber \\
	&+[\mu_S p (\uu\cdot \nabla) \EE(\rr)]\cdot \nabla_{\rr} f + [\mu_S^r p \uu \times \EE(\rr)]\cdot \rotop_{\uu} f\Big\},
\end{align}
where $\EE(\rr) = -\nabla \varphi(\rr)$ is the (Maxwell) electric field. Integrating by parts yields
\begin{align}
	\frac{\dd }{\dd t}f(\rr_j,\uu_j) =& \int \dd\rr \int\dd \uu \; f(\rr,\uu) \Big\{ -\sqrt{2D_S } \nabla_{\rr} \cdot(\boldsymbol{\xi}_j^S n_j)  - \sqrt{2D_S^r} \rotop_{\uu}\cdot(\boldsymbol{\zeta}_j^S n_j) + D_S\nabla^2_{\rr}n_j + D_S^r \rotop_{\uu}^2 n_j   \nonumber \\
	&-\nabla_{\rr} \cdot [n_j \mu_S p (\uu\cdot \nabla) \EE(\rr)] -  \rotop_{\uu}\cdot [n_j\mu_S^r p \uu \times \EE(\rr)] \Big\},
\end{align}
where we used an extension of integration by parts for the rotational gradient operator (see Section~\ref{app_rotop}). Since Eq.~\eqref{testfun} holds for any test function $f$, we deduce 
\begin{align}
	\partial_t n_j(\rr,\uu,t) =& -\sqrt{2D_S } \nabla_{\rr} \cdot(\boldsymbol{\xi}_j^S n_j) - \sqrt{2D_S^r} \rotop_{\uu}\cdot(\boldsymbol{\zeta}_j^S n_j) + D_S\nabla^2_{\rr}n_j + D_S^r \rotop_{\uu}^2 n_j   \nonumber\\
	&-\nabla_{\rr} \cdot [n_j \mu_S p (\uu\cdot \nabla) \EE(\rr)] -  \rotop_{\uu}\cdot [n_j\mu_S^r p \uu \times \EE(\rr)]
\end{align}
Summing for $j=1,\dots,N_S$ and using the usual tricks to rewrite the noise terms~\cite{Dean1996}, we get the equation for $n_S$:
\begin{align}
	\partial_t n_S(\rr,\uu,t) =&  D_S\nabla^2_{\rr}n_S + D_S^r \rotop_{\uu}^2 n_S   +\nabla_{\rr} \cdot [n_S \mu_S p (\uu\cdot \nabla) \nabla\varphi(\rr)] +  \rotop_{\uu}\cdot [n_j\mu_S^r p \uu \times \nabla \varphi(\rr)] \nonumber\\
	&+\nabla_{\rr} \cdot(\sqrt{2D_S n_S } \boldsymbol{\zeta}_S(\rr,\uu,t)) +\rotop_{\uu} \cdot(\sqrt{2D_S^r n_S } \boldsymbol{\zeta}_{S,r}(\rr,\uu,t)),
\end{align}
which is Eq.~\eqref{DK3} in the main text, and where $\boldsymbol{\zeta}_S(\rr,t)$  and $\boldsymbol{\zeta}_{S,r}(\rr,t)$ are independent unit white noises:
\begin{align}
	\langle \zeta_\alpha(\rr,\uu,t)\rangle &= 0 \\
	\langle\zeta_\alpha(\rr,\uu,t) \zeta_\beta(\rr',\uu',t')\rangle &= \delta_{\alpha\beta} \delta(\rr-\rr')\delta(\uu-\uu')\delta(t-t'),
\end{align}
for $\alpha,\beta=S$ or $S,r$.

\subsection{Some properties of the rotational gradient operator}
\label{app_rotop}

In this Section, we give a few properties of the rotational gradient operator $\rotop_{\uu}$. We recall its definition: $\rotop_{\uu} = \uu \times \nabla_{\uu}$, whose components can be conveniently rewritten as $(\mathcal{R}_{\uu})_i = \epsilon_{ijk} u_j \partial_{u_k}$, where $\epsilon_{ijk}$ is the Levi-Civita tensor. The operator $\rotop_{\uu}$ has the following properties:
\begin{itemize}
	\item Integration by parts: 
	\begin{equation}
		\int \dd \uu f(\uu) \rotop_{\uu}g (\uu) = -\int \dd \uu g(\uu) \rotop_{\uu} f(\uu),
	\end{equation}
	where the integrals are performed over the unit sphere.
	\item $\rotop_{\uu} \cdot \uu =0$ (since the basic properties of the Levi-Civita tensor imply $\varepsilon_{iji}=0)$).
	\item $\rotop_{\uu}^2 \uu = -2 \uu$ (this can be demonstrated using the following properties: $\varepsilon_{ijk}=-\varepsilon_{ikj}$ and $\varepsilon_{ijk}\varepsilon_{ijl} = 2 \delta_{kl}$).
\end{itemize}

\subsection{It\^o's lemma for orientational dynamics}
\label{app_Ito}

As a complement to the calculation detailed in the previous section, we give here details on It\^o's lemma for orientational dynamics. We consider a random vector $\nn$ whose modulus is constrained to be $|\nn|=1$. It is assumed to obey the following general stochastic differential equation:
\begin{equation}
	\frac{\dd \nn}{\dd t} = \boldsymbol{\omega}[\nn(t),t] \times \nn(t) + b[\nn(t),t] \boldsymbol{\zeta}(t) \times \nn(t) 
\end{equation}
where $\boldsymbol{\omega}[\nn(t),t]$ is the angular velocity, and the noise $\boldsymbol{\zeta}(t)$ is such that $\langle \zeta_i(t) \rangle=0$ and $\langle \zeta_i(t) \zeta_j(t') \rangle=\delta_{ij}\delta(t-t')$ (note that the fact the first time derivative of $\nn$ is written as a cross product between itself and an arbitrary vector ensures that its norm is constant).  We also define the Wiener process $\boldsymbol{Z}(t)$:
\begin{equation}
	\frac{\dd \boldsymbol{Z}}{\dd t} = \boldsymbol{\zeta}(t),
\end{equation}
which implies $\boldsymbol{Z}(t+\tau)-\boldsymbol{Z}(t) = \int_t^{t+\tau} \dd s \;  \boldsymbol{\zeta}(s)$. We then consider an arbitrary function of the random variable $\nn(t)$, defined as $Y(t) \equiv F[\nn(t),t]$. In the limit of a small time increment $\dd t$, we write:
\begin{align}
	&  Y(t + \dd t)-Y(t) \underset{\dd t \to 0}{=} \partial_{n_i} F[\nn(t)] (\boldsymbol{\omega}[\nn(t),t]\times \nn(t)  \dd t + \boldsymbol{\zeta}(t)\times \nn(t)  \dd t)_i \nonumber\\
	&+\frac{1}{2}  \partial_{n_i}\partial_{n_j}  F[\nn(t)](\boldsymbol{\omega}[\nn(t),t]\times \nn(t)  \dd t +  b[\nn(t),t] \boldsymbol{\zeta}(t)\times \nn(t)  \dd t)_i (\boldsymbol{\omega}[\nn(t),t]\times \nn(t)  \dd t +  b[\nn(t),t] \boldsymbol{\zeta(t)}\times \nn(t)  \dd t)_j + \dots \label{Yexp}
\end{align}

The first term is treated as follows. It is rewritten as:
\begin{align}
	\partial_{n_i} F[\nn(t)] (\boldsymbol{\omega}[\nn(t),t]\times \nn(t)  \dd t + \boldsymbol{\zeta}(t)\times \nn(t)  \dd t)_i &= \dd t [(\boldsymbol{\omega}[\nn(t),t]  +  b[\nn(t),t] \boldsymbol{\zeta}(t))\times \nn(t) ]  \cdot \nabla_{\nn} F[\nn(t)] \\
	&= \dd t (\boldsymbol{\omega}[\nn(t),t]  +  b[\nn(t),t] \boldsymbol{\zeta}(t)) \cdot \rotop_{\nn} F[\nn(t)] 
\end{align}
where we used the generic relation $ (\boldsymbol{A} \times \boldsymbol{B}) \cdot \boldsymbol{C}= \boldsymbol{A} \cdot (\boldsymbol{B} \times \boldsymbol{C})$ and the rotational gradient operator discussed in the previous section. The second term in Eq.~\eqref{Yexp} is  expanded in powers of $\dd t$, and contributions of order $\mathcal{O}(\dd t^2)$  are discarded. Since $\boldsymbol{\zeta}(t) \dd t = \dd \boldsymbol{Z}(t)$ is a Wiener process, we use the relation~\cite{Gardiner1985}
\begin{equation}
	\dd Z_i(t) \dd Z_j(t) = \delta_{ij}\dd t,
\end{equation}
and show that the second term in Eq.~\eqref{Yexp} reduces to
\begin{equation}
	\frac{1}{2} b[\nn(t),t]^2\partial_i \partial_j F[\nn(t)] (\boldsymbol{\zeta}(t)\times \nn(t) \dd t)_i(\boldsymbol{\zeta}(t)\times \nn(t) \dd t)_j =  \frac{1}{2} b[\nn(t),t]^2 \dd t \rotop_{\nn}^2 F[\nn(t)].
\end{equation}
Finally, we find that 
\begin{align}
	Y(t + \dd t)-Y(t) \underset{\dd t \to 0}{=} \dd t (\boldsymbol{\omega}[\nn(t),t]  +  b[\nn(t),t] \boldsymbol{\zeta}(t)) \times \rotop_{\nn} F[\nn(t)] +  \frac{1}{2} b[\nn(t),t]^2 \dd t \rotop_{\nn}^2 F[\nn(t)] + \mathcal{O}(\dd t^2),
\end{align}
or, equivalently,
\begin{equation}
	\frac{\partial}{\partial t} F[\nn(t)] = \boldsymbol{\omega}[\nn(t),t]\cdot \rotop_{\nn} F[\nn(t)]+ b[\nn(t),t] \boldsymbol{\zeta}(t) \cdot \rotop_{\nn} F[\nn(t)] + \frac{1}{2}b[\nn(t),t]^2 \rotop_{\nn}^2 F[\nn(t)]
\end{equation}
This is a reformulation of It\^o's lemma for the particular case where the random variable is a {unit} vector.

\subsection{SDFT equations and their linearization}

For completeness, we give here the SDFT equations satisfied by the densities of anions and cations (see Ref.~\cite{Dean1996} of~\cite{Demery2015a} for details on the derivation):
\begin{align}
\partial_t n_+ (\rr,t) &= D_I\nabla^2 n_+ + \mu_I e \nabla\cdot(n_+\nabla \varphi)+\nabla \cdot(\sqrt{2D_In_+} \boldsymbol{\zeta}_+), \\
\partial_t n_- (\rr,t) &= D_I\nabla^2 n_- - \mu_I e \nabla\cdot(n_-\nabla \varphi)+\nabla \cdot(\sqrt{2D_I n_-} \boldsymbol{\zeta}_-),
\end{align}
where $\boldsymbol{\zeta}_\pm$ are independent unit white noises, i.e. for $a,b=\pm$:
\begin{align}
    \langle \zeta_a(\rr,t) \rangle &= 0, \\
    \langle \zeta_a(\rr,t) \zeta_b(\rr',t') \rangle &= \delta_{ab}\delta(\rr-\rr')\delta(t-t'). 
\end{align}

From these equations, it is straightforward to deduce the evolution equations of the ionic charge density $\rho_I=ze(n_+-n_-)$ and the total number density $\mathcal{C} = n_++ n_-$, which read
\begin{eqnarray}
    \partial_t \rho_I(\rr,t) &=& D_I\nabla^2 \rho_I+\mu_I (ze)^2 \nabla \cdot (\mathcal{C} \nabla \varphi)+\nabla \cdot (ze\sqrt{2D_I\mathcal{C}} \boldsymbol{\zeta}_\rho), \\
    \partial_t \mathcal{C}(\rr,t) &=& D_I\nabla^2 \mathcal{C}+\mu_I  \nabla \cdot (\rho_I \nabla \varphi)+\nabla \cdot (\sqrt{2D_I\mathcal{C}} \boldsymbol{\zeta}_\mathcal{C}). 
\end{eqnarray}
This is completed by the equation satisfied by $n_S$, that we recall:
\begin{align}
&        \partial_t n_S(\rr,\uu,t) =  D_S\nabla^2_{\rr}n_S + D_S^r \rotop_{\uu}^2 n_S  +\nabla_{\rr} \cdot [n_S \mu_S p (\uu\cdot \nabla) \nabla\varphi(\rr)] +  \rotop_{\uu}\cdot [n_S\mu_S^r p \uu \times \nabla \varphi(\rr)] \nonumber\\
    &+\nabla_{\rr} \cdot(\sqrt{2D_S n_S } \boldsymbol{\zeta}_S(\rr,\uu,t)) +\rotop_{\uu} \cdot(\sqrt{2D_S^r n_S } \boldsymbol{\zeta}_{S,r}(\rr,\uu,t)).  \label{DK3_supp}
\end{align}

The linearization mentioned in the main text consists in assuming that the ionic densities remain close to a constant, uniform value $C_I$: $n_\pm = C_I + \delta n_\pm$ (in such a way that $\rho_I=ze(\delta n_+-\delta n_-)$ is of order $1$, and $\mathcal{C}=2C_I+c$). Similarly, the density of solvent molecules is linearized as $n_S = C_S/4\pi + c_S$. The linear equation by $\rho_I$, $c$ and $c_S$ read
\begin{eqnarray}
 \partial_t \rho_I(\rr,t) &=& D_I\nabla^2 \rho_I+2\mu (ze)^2 C_I \nabla^2\varphi+ze\sqrt{4DC_I} {\eta}_{\rho_I}(\rr,t), \label{lineq_rho}  \\
  \partial_t c(\rr,t) &=& D_I\nabla^2 c+ \sqrt{4DC_I} {\eta}_c(\rr,t), \label{lineq_c}  \\
 \partial_t c_S(\rr,\uu,t) &=& D_S\nabla^2 c_S + D_S^r \rotop^2 c_S + \mu p \frac{C_S}{4\pi} \nabla \cdot [(\uu\cdot \nabla)\nabla \varphi]   + \mu_S^r p \frac{C_S}{4\pi} \rotop \cdot[\uu\times \nabla\varphi] \nonumber\\
    &&+\sqrt{2D_S} \sqrt{\frac{C_S}{4\pi}} \eta_S (\rr,\uu,t)+\sqrt{2D_S^r} \sqrt{\frac{C_S}{4\pi}} \eta_{S,r}(\rr,\uu,t)
    \label{cs_lin}
\end{eqnarray}
The noises $\eta_\rho$, $\eta_S$ and $\eta_S^r$ are uncorrelated, have zero average, and  correlations: 
\begin{align}
    \langle \eta_{\rho_I}(\rr,t) \eta_{\rho_I}(\rr',t')\rangle&=-\delta(t-t') \nabla_{\rr}^2\delta(\rr-\rr')\\
    \langle \eta_S(\rr,\uu,t) \eta_S(\rr',\uu',t')\rangle&=-\delta(t-t') \delta(\uu-\uu')\nabla_{\rr}^2\delta(\rr-\rr')\\
    \langle \eta_S^r(\rr,\uu,t) \eta_S^r(\rr',\uu',t')\rangle&=-\delta(t-t') \delta(\rr-\rr') \rotop^2\delta(\uu-\uu')
\end{align}

\subsection{Case of an implicit solvent}

For ions in a implicit solvent, represented by a continuous medium of permittivity $\varepsilon_0 \varepsilon_\mathrm{w}$ (where $\varepsilon_\mathrm{w}$ is the relative permittivity of pure water), Poisson's equation simply reads:
\begin{equation}
    -\nabla^2\varphi = \frac{\rho_I^\text{imp}}{\varepsilon_0 \varepsilon_\mathrm{w}}
\end{equation}
The linear equation satisfied by $\rho_I^\text{imp}$ [Eq.~\eqref{lineq_rho}] is then closed, and takes the simple form:
\begin{equation}
\partial_t \rho^\text{imp}_I(\rr,t) = D_I\nabla^2 \rho^\text{imp}_I-D_I\kappa_\text{imp}^2 \rho^\text{imp}_I+ze\sqrt{4DC_I} {\eta}_{\rho_I}, 
\end{equation}
with $\kappa_\text{imp} = 2(ze)^2C_I/\varepsilon_0 \varepsilon_\mathrm{w} \kB T$. In Fourier space, the solution reads:
\begin{equation}
\tilde \rho^\text{imp}_I(\qq,t) = ze\sqrt{4DC_I} \int_{-\infty}^{t} \dd t'\; {\eta}_{\rho_I}(\qq,t') \ex{-D_I(q^2+\kappa_\text{imp}^2 )(t-t')}.
\end{equation}
Computing the structure factor $F_{II,\text{imp}}(q,t)=\frac{1}{2 N_I} \langle \tilde \rho^\text{imp}_I(\qq,t) \tilde\rho^\text{imp}_I(-\qq,0)\rangle$ yields, for $t>0$:
\begin{equation}
    F_\text{imp}(q,t) = \frac{(ze)^2 q^2}{q^2+\kappa_\text{imp}^2} \ex{-D_I(q^2+\kappa_\text{imp}^2 )t}
\end{equation}
which is the expression given in the main text.

\subsection{Polarization field}
\label{sec_polarization}

\subsubsection{Equation for the polarization field}

An equation for the polarisation field $\PP$ can be obtained by multiplying Eq.~\eqref{cs_lin} by $\uu$ and integrating over all orientations $\uu$:
\begin{align}
    \partial_t \PP(\rr,t) =& D_S \nabla^2 \PP -2D_S^r \PP +\frac{1}{3}p^2 C_S \nabla(\mu_S \nabla^2 \varphi-2\mu_R \varphi) + \boldsymbol{\Xi} (\rr,t) ,
    \label{lineq_P2_supp}
\end{align}
with 
\begin{align}
    \langle \Xi_k(\rr,t) \Xi_l(\rr',t') \rangle= \frac{2 C_S p}{3} \delta_{kl} \delta (t-t') (-D_S \nabla^2+2D_S^r) \delta (\rr-\rr').
\end{align}
This is Eq.~\eqref{lineq_P2} in the main text. In order to get this result, we relied on the following intermediate steps: (i) the second term in the rhs of Eq.~\eqref{cs_lin} is treated by using the relation $\rotop^2 \uu = -2 \uu$; (ii) the third and fourth terms in the rhs of  Eq.~\eqref{cs_lin} are computed by using the relation $\int \dd \uu \; u_i u_j = \frac{4\pi}{3} \delta_{ij}$. Importantly, we note that multiplying Eq.~\eqref{cs_lin} by $\uu$ and integrating over $\uu$ does not yield the unclosed usual hierarchy of equations that appear in `moment expansions' of active matter theory~\cite{Ahmadi2006,Brotto2013,Golestanian2019}: charge conservation imposes intrinsically different symmetries in the equations of motion.

\subsubsection{Charge polarization}

 At this stage, it is useful to recall that, within the dipolar approximation, the electrostatic potential $\varphi$ only depends on the divergence of $\PP$, i.e. on the longitudinal part of the field $\PP$ in Fourier space. We then need an equation for $\rho_S= -\nabla \cdot \PP$ (or in Fourier space, $\tilde\rho_S = -\ii\qq\cdot \tilde\PP$) instead of $\PP$. To this end, we take the Fourier transform of Eq.~\eqref{lineq_P2} for a given component $i$:
\begin{align}
    \partial_t \tilde P_i(\qq,t) =& -D_S q^2 \tilde P_i -2D_S^r \tilde P_i  +\frac{1}{3}p^2 C_S \ii q_i(- q^2\mu_S \tilde\varphi-2\mu_S^r \tilde \varphi) +\tilde\Xi_i(\qq,t) 
\end{align}
Multiplying by $- \ii q_i$ and summing over all $i$ yields
\begin{align} 
    \partial_t \tilde \rho_S(\qq,t) =& -D_S q^2\tilde \rho_S -2D_S^r\tilde \rho_S  +\frac{1}{3}p^2 C_S  q^2(- q^2\mu_S \tilde\varphi-2\mu_S^r \tilde \varphi) +{\Xi}_S(\qq,t) ,
\end{align}
with
\begin{equation}
    \langle \Xi_S (\qq,t) \Xi_S(\qq',t') \rangle = \frac{2C_S}{3} p^2 q^2 (D_S q^2 + 2 D_S^r)   (2\pi)^d \delta (\qq+\qq') \delta(t-t').
  \end{equation}
We finally recall that
\begin{equation}
\tilde\varphi(\qq,t)=\tilde{\mathcal{G}}(\qq) [\tilde \rho_I(\qq,t)+\tilde \rho_S(\qq,t)] = \frac{1}{\varepsilon_0 q^2} [\tilde \rho_I(\qq,t)+\tilde \rho_S(\qq,t)] 
\end{equation}
We then get:
\begin{align}
    \partial_t \tilde \rho_S(\qq,t) &= -D_S q^2\tilde \rho_S -2D_S^r\tilde \rho_S  -\frac{p^2 C_S}{3\varepsilon_0}  ( q^2\mu_S +2\mu_S^r ) (\tilde\rho_I+\tilde\rho_S) +{\Xi}_S(\qq,t) .
\end{align}
This yields the second equation of the linear system given in the main text as Eq.~\eqref{eq_main_result}.

\subsubsection{Transverse and longitudinal components of the polarization field}
\label{P_components}

Finally, for completeness, we also derive the evolution equation for the transverse component of the polarisation field $\PP$. In Fourier space, we write the following decomposition:
\begin{equation}
\tilde P_i(\qq,t) = \underbrace{ \frac{q_i q_j}{q^2}  \tilde P_j(\qq,t)}_{\equiv \tilde P_{L,i}(\qq,t)} + \underbrace{  \left( \delta_{ij} - \frac{q_i q_j}{q^2}  \right)  \tilde P_j(\qq,t)}_{\equiv \tilde P_{T,i}(\qq,t)}
\end{equation}
It is clear that $ \tilde P_{L,i}(\qq,t)= \ii\frac{q_i}{q^2}\tilde \rho_S$. We deduce the evolution equations for the longitudinal and transverse components of the polarisation field:
\begin{align}
    \partial_t \tilde \PP_L(\qq,t) &= -D_S q^2\tilde \PP_{L} -2D_S^r\tilde \PP_L  -D_S\kappa_S(q)^2 \left( \ii\frac{\qq}{q^2} \tilde\rho_I+ \tilde \PP_{L} \right) +\boldsymbol{\Xi}_L(\qq,t) \\
        \partial_t \tilde \PP_{T}(\qq,t) &= -(D_S q^2+2D_S^r)\tilde \PP_T   +\boldsymbol{\Xi}_T(\qq,t) 
\end{align}
where the noises $\boldsymbol{\Xi}_L(\qq,t)$ and $\boldsymbol{\Xi}_T(\qq,t)$ are uncorrelated and have respective variances:
\begin{align}
\langle \Xi_{L,i}(\qq,t) \Xi_{L,j}(\qq',t')\rangle &=  2  \frac{q_i q_j}{q^2} D_S \kappa_S(q)^2 \varepsilon_0\kB T (2\pi)^d   \delta(\qq+\qq')\delta(t-t'),\\
\langle \Xi_{T,i}(\qq,t) \Xi_{T,j}(\qq',t')\rangle &=  2 \left( \delta_{ij} -\frac{q_i q_j}{q^2} \right)D_S \kappa_S(q)^2 \varepsilon_0\kB T (2\pi)^d   \delta(\qq+\qq')\delta(t-t').
\end{align}
where $\kappa_S(q)^2$ is the screening length associated to solvent defined in the main text.

\section{Free energy associated to pure water}

In the situation where there are no ions in the system, the polarization field obeys the (closed) equation:
\begin{equation}
    \partial_t \PP(\rr,t) = D_S \nabla^2 \PP -2D_S^r \PP +\frac{1}{3}p^2 C_S \nabla\left[\frac{\mu_S}{\varepsilon_0} \nabla\cdot \PP+2 \mu_S^r \int\dd \rr' \mathcal{G}(\rr-\rr')\nabla\cdot \PP(\rr')\right] + \boldsymbol{\Xi} (\rr,t) 
\end{equation}
The goal is to rewrite this equation under the form:
\begin{equation}
     \partial_t \PP(\rr,t) = -\mu_S \frac{\delta \mathcal{F}[\PP]}{\delta\PP(\rr,t)}  + \boldsymbol{\Xi} (\rr,t)
\end{equation}
where $\mathcal{F}$ is a free energy to be determined. To this end, we use the following useful functional derivatives:
\begin{align}
&\frac{\delta}{\delta P_i(\rr)}\int \dd \rr' \;  \PP(\rr')^2 = 2P_i(\rr) \\
&\frac{\delta}{\delta P_i(\rr)}\int \dd \rr' \;  (\nabla\cdot \PP(\rr'))^2 = -2 \partial_i \nabla \cdot \PP(\rr) \\
&\frac{\delta}{\delta P_i(\rr)}\int \dd \rr' \;   [\nabla(\nabla\cdot \PP(\rr'))]^2= 2\partial_i \nabla^2(\nabla\cdot \PP(\rr) ) \\
&\frac{\delta}{\delta P_i(\rr)}\int \dd \rr' \int \dd \rr''\; \mathcal{G}(\rr'-\rr'') \nabla\cdot\PP(\rr')  \nabla\cdot\PP(\rr'') =-2\partial_i  \int \dd \rr'\; \mathcal{G}(\rr-\rr') \nabla\cdot\PP(\rr')   \\
&\frac{\delta}{\delta P_i(\rr)}\int \dd \rr' \; (\partial_kP_j(\rr'))(\partial_kP_j(\rr')) =-2\nabla^2 P_i(\rr) 
\end{align}
We then get the following free energy:
\begin{align}
\mathcal{F}[\PP] = \kB T\int \dd \rr' \Bigg\{& \frac{1}{2} (\partial_kP_j(\rr'))(\partial_kP_j(\rr')) +\frac{1}{a^2}  \PP(\rr')^2 + \frac{1}{4}a^2\kappa_{S}(0)^2   (\nabla\cdot \PP(\rr'))^2  \nonumber\\
 &+ \frac{1}{2} \varepsilon_0\kappa_S(0)^2 \int \dd \rr''\; \mathcal{G}(\rr'-\rr'')   \nabla\cdot\PP(\rr'') \Bigg\}
\label{ourHamiltonian}
\end{align}
where the first line corresponds to `local' term and the second one to non-local terms. This is to be compared to the Hamiltonian that was proposed on phenomenological grounds, or through symmetry considerations~\cite{Berthoumieux2015, Berthoumieux2018, Vatin2021}:
\begin{eqnarray}
\mathcal{H}[\PP] &=&\frac{1}{2\varepsilon_0} \int \dd \rr' \left\{ K   \PP(\rr')^2 + \kappa_l (\nabla\cdot \PP(\rr'))^2   +\alpha(\nabla(\nabla\cdot \PP(\rr)))^2   \right\}  \nonumber\\
&&+ \frac{1}{2\varepsilon_0} \int \dd \rr' \int \dd \rr''\; \frac{ \nabla\cdot\PP(\rr')  \nabla\cdot\PP(\rr'') }{4\pi|\rr-\rr'|},
\label{HamiltonianHelene}
\end{eqnarray}
where the Landau coefficients $K$, $\kappa_l$ and $\alpha$ are typically estimated by comparison with response functions measured experimentally or from molecular simulations.

 When comparing the two free energies or Hamiltonians given by Eqs.~\eqref{ourHamiltonian} and~\eqref{HamiltonianHelene}, we  make two observations: (i)~our free energy functional contains a term proportional to $ (\partial_kP_j(\rr'))(\partial_kP_j(\rr'))$, which accounts for the translational diffusion of the solvent molecules. This term is  absent in the theories from Refs.~\cite{Maggs2006,Berthoumieux2015,Berthoumieux2018,Blossey2022,Blossey2022a,Becker2023}, who typically consider the diffusion of the solvent density and that of its polarization in different functionals (see e.g. Eq. (1) in Ref.~\cite{Berthoumieux2018}); (ii)~the last term in Eq.~\eqref{HamiltonianHelene} is typically obtained by Taylor expansion arguments (see e.g. Ref.~\cite{Maggs2006}) and does not appear in our derivations.

\section{Explicit solution of the coupled equations for $\tilde \rho_I$ and $\tilde \rho_S$}

\subsection{General solution}

We start from the main result of the Letter, given as Eq.~\eqref{eq_main_result}. Its solution is formally written as
\begin{equation}
    \begin{pmatrix}
        \tilde \rho_I(\qq,t) \\
        \tilde \rho_S(\qq,t)
    \end{pmatrix} = \int_{-\infty}^t \dd s \; \boldsymbol{\mathcal{M}}(t-s) 
    \begin{pmatrix}
        \Xi_I(\qq,s) \\
        \Xi_S (\qq,s)
    \end{pmatrix},
\end{equation}
where $ \boldsymbol{\mathcal{M}}(t)$ is the matrix exponential $\ex{-t \mathbf{M}}$:
\begin{equation}
    \boldsymbol{\mathcal{M}}(t) = \mathbf{c}^{(+)} \ex{-\lambda_+ t} + \mathbf{c}^{(-)} \ex{-\lambda_- t},
\end{equation}
with the matrices
\begin{equation}
    \mathbf{c}^{(\pm)} = \frac{1}{2s} \begin{pmatrix}
        \pm M_{11} \mp M_{22}+s & \pm 2 M_{12 } \\
        \pm 2 M_{21} & \mp M_{11} \pm M_{22}+s
    \end{pmatrix},
    \label{cpm}
\end{equation}
where $s = \sqrt{(M_{11}-M_{22})^2+4M_{12}M_{21}}$, and where we define the eigenvalues of $\mathbf{M}$:
\begin{equation}
    \lambda_\pm \equiv \frac{M_{11}+M_{22}}{2} \pm \frac{1}{2}\sqrt{(M_{11}-M_{22})^2+4M_{12}M_{21}}.
    \label{lambda_def}
\end{equation}

\subsection{Computing the dynamic charge structure factors}

The dynamic charge structure factors can be computed explicitly in terms of the elements of the matrices $\mathbf{c}^{(\pm)}$ and of the eigenvalues $\lambda_\pm$, i.e. in terms of all the parameters of the problem. We give as an example the two-point, two-time correlation function of $\tilde \rho_I$, which reads
\begin{align}
    \langle \tilde \rho_I(\qq,t) \tilde \rho_I(\qq',t') \rangle = &4DC_I(ze)^2(2\pi)^d q^2 \delta(\qq+\qq') \int_{-\infty}^{t'} \dd s' \mathcal{M}_{11}(\qq,t-s')\mathcal{M}_{11}(\qq,t'-s') \nonumber\\
    &+\frac{2 C_S}{3} p^2q^2 (2\pi)^d (D_S q^2+2D_S^r) \delta(\qq+\qq') \int_{-\infty}^{t'} \dd s' \mathcal{M}_{12}(\qq,t-s')\mathcal{M}_{12}(\qq,t'-s')
\end{align}
We assumed $t>t'$ with no loss of generality, and we used the fact that the entries $\mathcal{M}_{\alpha\beta}$ are even functions of $\qq$. We then use the following relation:
\begin{equation}
    \int_{-\infty}^{t'} \dd s' \mathcal{M}_{\alpha\beta}(\qq,t-s')\mathcal{M}_{\gamma\delta}(\qq',t'-s')  = \sum_{\epsilon=\pm 1} C^{(\epsilon)}_{\alpha\beta,\gamma\delta}(q)\ex{-\lambda_{\epsilon}(q) |t-t'|}, 
\end{equation}
with the fourth rank tensor:
\begin{equation}
\label{C_abcd}
C^{(\epsilon)}_{\alpha\beta,\gamma\delta}(q) = \frac{c_{\alpha\beta}^{(\epsilon)} \lambda_{-\epsilon}[c_{\gamma\delta}^{(\epsilon)} \lambda_{-\epsilon} + \lambda_\epsilon(2c_{\gamma\delta}^{(-\epsilon)}+c_{\gamma\delta}^{(\epsilon)})]}{2\lambda_1 \lambda_{-1}(\lambda_1 +\lambda_{-1})}
\end{equation}
being homogeneous to a time. We get, for arbitrary $t,t'$:
\begin{equation}
     \langle \tilde \rho_I(\qq,t) \tilde \rho_I(\qq',t') \rangle = (2\pi)^d \delta(\qq+\qq') \sum_{\epsilon =\pm 1} \ex{-\lambda_{\epsilon}|t-t'|}\left[4 DC_I (ze)^2 q^2  C_{11,11}^{(\epsilon)}(q) + \frac{2 C_S}{3} p^2q^2(D_S q^2+2D_S^r)C_{12,12}^{(\epsilon)}(q)\right].
\end{equation}
Using the expressions of the inverse screening lengths $\kappa$ and $\kappa_S$, and choosing $t'=0$ with no loss of generality, we get 
\begin{equation}
     \langle \tilde \rho_I(\qq,t) \tilde \rho_I(\qq',0) \rangle = (2\pi)^d \delta(\qq+\qq') 2q^2\varepsilon_0\kB T \sum_{\epsilon =\pm 1} \ex{-\lambda_{\epsilon}|t|}\left[  D \kappa_I^2  C_{11,11}^{(\epsilon)}(\qq) + D_S \kappa_S^2(q) C_{12,12}^{(\epsilon)}(\qq)\right] .
\end{equation}
For $\alpha,\beta\in\{I,S\}$, we get the general expression:
\begin{equation}
     \langle \tilde \rho_\alpha(\qq,t) \tilde \rho_\beta(\qq',0) \rangle = (2\pi)^d \delta(\qq+\qq') 2q^2\varepsilon_0 \kB T \sum_{\substack{\gamma\in\{S,I\}   \\ \epsilon =\pm1   }} D_\gamma \kappa_\gamma^2 C_{\alpha\gamma,\beta\gamma}^{(\epsilon)} \ex{-\lambda_\epsilon |t|}.
\end{equation}

We recall the usual definition of (partial) intermediate scattering functions, for a finite system (for instance a box of volume $V$ with periodic boundary conditions, and containing $\mathcal{N}_I=2 N_I $ ions and $\mathcal{N}_S= N_S$ solvent molecules):
\begin{equation} 
F_{\alpha\beta}(q,t) = \frac{1}{\sqrt{\mathcal{N}_\alpha\mathcal{N}_\beta}} \langle \tilde \rho_\alpha(\qq,t) \rho_\beta(-\qq,0)  \rangle .
\end{equation}
In infinite space, we demonstrated that $     \langle \tilde \rho_\alpha(\qq,t) \tilde \rho_\beta(\qq',0) \rangle = (2\pi)^d \delta(\qq+\qq') f_{\alpha\beta}(q,t)$ with 
\begin{equation}
f_{\alpha\beta }(q,t)=  2q^2\varepsilon_0 \kB T \sum_{\substack{\gamma\in\{S,I\}   \\ \epsilon =\pm1   }} D_\gamma \kappa_\gamma^2 C_{\alpha\gamma,\beta\gamma}^{(\epsilon)} \ex{-\lambda_\epsilon |t|}
\label{def_f}
\end{equation}
In a periodic system, the wave vectors $\qq$ are actually discrete, and the Fourier transform has a different expression.  We can map our results onto a discrete periodic system with the replacement $(2\pi)^d \delta(\qq+\qq') \to V \delta_{\qq,-\qq'}$. Therefore, it is clear that 
\begin{equation} 
F_{\alpha\beta}(q,t) = \frac{V}{\sqrt{\mathcal{N}_\alpha\mathcal{N}_\beta}}f_{\alpha\beta}(q,t) .
\end{equation}
In conclusion, we get the following expressions for the ISFs:
\begin{align}
&F_{II}(q,t) = \frac{q^2\varepsilon_0 \kB T }{C_I}   \sum_{\substack{\gamma\in\{S,I\}   \\ \epsilon =\pm1   }} D_\gamma \kappa_\gamma^2 C_{I\gamma,I\gamma}^{(\epsilon)} \ex{-\lambda_\epsilon |t|}, \\
&F_{SS}(q,t) = \frac{ 2q^2\varepsilon_0 \kB T }{C_S}  \sum_{\substack{\gamma\in\{S,I\}   \\ \epsilon =\pm1   }} D_\gamma \kappa_\gamma^2 C_{S\gamma,S\gamma}^{(\epsilon)} \ex{-\lambda_\epsilon |t|}, \\
&F_{SI}(q,t) =F_{IS}(q,t) = \frac{ 2q^2\varepsilon_0 \kB T }{\sqrt{2 C_S C_I}}  \sum_{\substack{\gamma\in\{S,I\}   \\ \epsilon =\pm1   }} D_\gamma \kappa_\gamma^2 C_{I\gamma,S\gamma}^{(\epsilon)} \ex{-\lambda_\epsilon |t|} .
\end{align}
This is the result given as Eq.~\eqref{ISF_full} in the main text.

\section{Dielectric properties}

\subsection{Susceptibilities}

When the electrolyte is submitted to a constant electric field $\EE_0$, it will show a macroscopic polarization ${\PP}$ and current ${\boldsymbol{j}}$, that are related to the microscopic polarization and current: 
\begin{eqnarray}
 \PP(\rr,t) &=& \sum_{j=1}^{N_s} p\uu_j(r) \delta(\rr-\rr_j^S(t)) \\
 \jj (\rr,t) &=& \sum_{i=1}^{N_I} ze \vv_i^+(t) \delta(\rr-\rr_i^+(t)) + \sum_{i=1}^{N_I} (-ze) \vv_a^-(t) \delta(\rr-\rr_i^-(t)) \label{def_current}
\end{eqnarray}
Within linear response theory, the macroscopic observables ${\PP}$ and ${\boldsymbol{j}}$ are related to the external field through the relations in Fourier space:
\begin{eqnarray}
 \PP(\qq,\omega) &=& \boldsymbol{\chi}_{\PP}(\qq,\omega)\cdot \EE_0(\qq,\omega) \\
 \jj(\qq,\omega) &=& \boldsymbol{\chi}_{\jj}(\qq,\omega)\cdot \EE_0(\qq,\omega) 
\end{eqnarray}
where $\boldsymbol{\chi}_{\PP}$ and  $\boldsymbol{\chi}_{\jj}$ are the second or external susceptibilities~\cite{Hansen1986}.\\

In what follows, each vector field $\boldsymbol{A}$ ($\boldsymbol{A}=\PP$ or $\jj$) is decomposed into a longitudinal and transverse part as:
\begin{equation} 
A_i = \underbrace{\hat q_i \hat q_j A_j}_{\equiv A_{L,i}} + \underbrace{\left( \delta_{ij }-\hat q_i \hat q_j \right) A_j}_{\equiv A_{T,i}} ,
\end{equation}
where we define the unit vector $\hat{\qq}=\qq/q$. In other words, the longitudinal and transverse parts of the fields can be compute as: $\boldsymbol{A}_L = (\hat{\qq}\cdot\boldsymbol{A})\qq$ and $\boldsymbol{A}_T = \boldsymbol{A} -  (\hat{\qq}\cdot\boldsymbol{A})\hat{\qq}$. Similarly, the susceptibility tensors are decomposed as:
\begin{equation}
    \boldsymbol{\chi}_{\boldsymbol{A}}(\qq,\omega) = \chi _{\boldsymbol{A},L}(\qq,\omega)\hat{\qq}\hat{\qq} + \chi _{\boldsymbol{A},T}(\qq,\omega) (\mathbf{1}-\hat{\qq}\hat{\qq}),
\end{equation}
where $\mathbf{1}$ is the unit tensor.\\

The susceptibilities can be computed from the correlation functions of the microscopic quantities using the relations~\cite{Ladanyi1999, Caillol1987,Madden1984,Caillol1986}:
\begin{align}
\chi_{\boldsymbol{P},L}(\qq,\omega) &= \frac{1}{V\varepsilon_0 \kB T} \left[ \langle \PP_L(\qq,0) \cdot \PP_L(-\qq,0) \rangle + \int_0^\infty \dd t \; \ex{\ii \omega t} \left\{  \ii\omega \langle \PP_L(\qq,t) \cdot \PP_L(-\qq,0) \rangle+\langle \PP_L(\qq,t) \cdot \jj_L(-\qq,0) \rangle \right\}  \right], \label{suscep1} \\
\chi_{\boldsymbol{P},T}(\qq,\omega) &= \frac{1}{2V\varepsilon_0 \kB T} \left[ \langle \PP_T(\qq,0) \cdot \PP_T(-\qq,0) \rangle + \int_0^\infty \dd t \; \ex{\ii \omega t} \left\{  \ii\omega \langle \PP_T(\qq,t) \cdot \PP_T(-\qq,0) \rangle+\langle \PP_T(\qq,t) \cdot \PP_T(-\qq,0) \rangle \right\}  \right], \\
\chi_{\boldsymbol{j},L}(\qq,\omega) &= \frac{1}{V \kB T}  \int_0^\infty \dd t \; \ex{\ii \omega t} \left\{  \ii\omega \langle \jj_L(\qq,t) \cdot \PP_L(-\qq,0) \rangle+\langle \jj_L(\qq,t) \cdot \jj_L(-\qq,0) \rangle \right\} ,  \\ 
\chi_{\boldsymbol{j},T}(\qq,\omega) &= \frac{1}{2V \kB T}  \int_0^\infty \dd t \; \ex{\ii \omega t} \left\{  \ii\omega \langle \jj_T(\qq,t) \cdot \PP_T(-\qq,0) \rangle+\langle \jj_T(\qq,t) \cdot \jj_T(-\qq,0) \rangle \right\}.    \label{suscep4}
\end{align}

In the main text, we defined the total susceptibility $\boldsymbol{\chi} = \boldsymbol{\chi}_{\PP} + \frac{\ii}{\omega \varepsilon_0} \boldsymbol{\chi}_{\jj}$, and we compute its longitudinal and transverse components, whose expression are deduced from Eqs.~\eqref{suscep1}-\eqref{suscep4} .

\subsection{Microscopic correlation functions}

We now proceed to compute the microscopic correlation functions that appear in Eqs.~\eqref{suscep1}-\eqref{suscep4}. To this end, we need to give the equations satisfied by the longitudinal and transverse parts of the polarisation and of the ionic currents.

\subsubsection{Polarisation}

We use the results from Section~\ref{P_components} to write:
\begin{align}
\partial_t \PP_L &= -D_S q^2 \PP_L-2D_S^r \PP_L -D_S\kappa_S(q)^2  \left(\ii \frac{\qq}{q^2}\tilde\rho_I + \PP_L \right)+ \boldsymbol{\Xi}^P_L(\qq,t), \label{PL_final} \\
\partial_t \PP_T &=-(D_Sq^2 + 2D_S^r) \PP_T+\boldsymbol{\Xi}^P_T(\qq,t), \label{PT_final}
\end{align}
with 
\begin{align}
\langle \Xi^P_{L,i}(\qq,t)\Xi^P_{L,j}(\qq',t')\rangle &=2  \frac{q_i q_j}{q^2} \varepsilon \kB T D_s \kappa_S(q)^2 (2\pi)^d \delta(\qq+\qq') \delta(t-t'),  \\
\langle \Xi^P_{L,i}(\qq,t)\Xi^P_{L,j}(\qq',t')\rangle &=  2\left( \delta_{ij}- \frac{q_i q_j}{q^2}\right) \varepsilon \kB T D_s \kappa_S(q)^2 (2\pi)^d \delta(\qq+\qq') \delta(t-t'),  \\
\langle \Xi^P_{L,i}(\qq,t)\Xi^P_{T,j}(\qq',t')\rangle &= 0 .
\end{align}

\subsubsection{Current}

In order to compute the ionic current $\jj(\rr,t)$ defined in Eq.~\eqref{def_current}, we first define the current associated to anions and cations:
\begin{equation}
\jj_\pm(\rr,t) = \sum_{a=1}^{N_I} \vv_a^\pm(t) \delta(\rr-\rr_a^\pm(t)),
\end{equation}
in such a way that $\jj=ze(\jj_+-\jj_-)$. We write the equations satisfied by $\vv_a^\pm$ and $\rr_a^\pm$ in the underdamped limit:
\begin{align}
\frac{\dd \rr_a^\pm}{\dd t} &= \vv_a^\pm(t), \\
m_I \frac{\dd \vv_a^\pm}{\dd t} &= -\zeta_\pm \vv_a^\pm \pm ze \EE(\rr_a^\pm(t))  + \sqrt{2\zeta_\pm \kB T} \boldsymbol{\Xi}^\pm_a(t),
\end{align}
where $m_I$ is the mass of the ions. The standard framework proposed by Dean, which holds for overdamped dynamics, was extended to underdamped dynamics by Nakamura and Yoshimori~\cite{Nakamura2009,Demery2015,Das2013}. We apply it to derive the following equations for $n_\pm$ and $\jj_\pm$:
\begin{eqnarray}
\partial_t n_\pm &=& -\nabla \cdot \jj_\pm \\
\partial_t \jj_\pm &=& -\frac{1}{\tau_\pm}\jj_\pm  -\frac{D}{\tau_\pm}\nabla n_\pm-\frac{\mu}{\tau_\pm} n_\pm (\pm ze )\nabla \varphi  -\nabla \cdot \left( \frac{\jj_\pm\jj_\pm^T}{n_\pm}\right) + \frac{\sqrt{2Dn_\pm}}{\tau\pm} \boldsymbol{\zeta}_\pm .
\end{eqnarray}
with the inertial times $\tau_\pm=m_\pm/ \zeta_\pm$. Although the expression of $\jj_\pm$ is therefore quite complicated in the general case, it takes a much simpler expression in the overdamped limit, i.e. when $\tau_\pm \to 0$:
\begin{equation}
\jj_\pm =  -  D\nabla n_\pm -{\mu} n_\pm (\pm ze )\nabla \varphi  + \sqrt{2Dn_\pm}\boldsymbol{\zeta}_\pm .
\end{equation}
It is straightforward to deduce the charge current in the linearized limit:
\begin{equation}
\jj = ze(\jj_+ - \jj_-)= -D\nabla \rho_I -2\mu (ze)^2 C_I\nabla\varphi+ \sqrt{4DC_I} \boldsymbol{\xi}(\rr,t).
\end{equation}
We finally compute the Fourier transform of $\jj$ and project it onto the longitudinal and transverse directions to get:
\begin{align}
\jj_L(\qq,t) &= -\ii \qq D\tilde\rho_I - D\kappa_I^2 \frac{\ii \qq}{q^2} \tilde\rho_I  - D\kappa_I^2\PP_L+\boldsymbol{\Xi}^j_L(\qq,t), \label{jL_final}\\
\jj_T(\qq,t) &=\boldsymbol{\Xi}^j_T(\qq,t), \label{jT_final}
\end{align}
with
\begin{align}
\langle \Xi^j_{L,i}(\qq,t)\Xi^j_{L,j}(\qq',t')\rangle &=2  \frac{q_i q_j}{q^2} \varepsilon \kB T D_I \kappa_I^2 (2\pi)^d \delta(\qq+\qq') \delta(t-t'),  \\
\langle \Xi^j_{T,i}(\qq,t)\Xi^j_{T,j}(\qq',t')\rangle &=  2\left( \delta_{ij}- \frac{q_i q_j}{q^2}\right) \varepsilon \kB T D_I\kappa_I^2 (2\pi)^d \delta(\qq+\qq') \delta(t-t') , \\
\langle \Xi^j_{L,i}(\qq,t)\Xi^j_{T,j}(\qq',t')\rangle &= 0 .
\end{align}

\subsubsection{Final equations for the correlation functions}

Putting together the results from Eqs.~\eqref{PL_final},~\eqref{PT_final},~\eqref{jL_final},  and~\eqref{jT_final}, we get (see Eq.~\eqref{def_f} for the definition of the functions $f_{\alpha\beta}$):
\begin{itemize}
\item Longitudinal contributions:
\begin{eqnarray}
\frac{1}{V}\langle \PP_L(\qq,t)\cdot\PP_L(-\qq,0)  \rangle &=& \frac{1}{q^2} f_{SS}(q,t) \\
\frac{1}{V}\langle \PP_L(\qq,t)\cdot\jj_L(-\qq,0)  \rangle &=& -D\left( 1+\frac{\kappa_I^2}{q^2}  \right) f_{SI}(q,t) -D\frac{\kappa_I^2}{q^2}  f_{SS}(q,t)   \\
\frac{1}{V}\langle \jj_L(\qq,t)\cdot\PP_L(-\qq,0)  \rangle &=& -D\left( 1+\frac{\kappa_I^2}{q^2}  \right) f_{IS}(q,t) -D\frac{\kappa_I^2}{q^2}  f_{SS}(q,t)   \\
\frac{1}{V}\langle \jj_L(\qq,t)\cdot\jj_L(-\qq,0)  \rangle &=& D^2 q^2\left( 1+\frac{\kappa_I^2}{q^2}  \right)^2 f_{II}(q,t)+D^2 q^2\left(\frac{\kappa_I^2}{q^2} \right)^2 f_{SS}(q,t)   \nonumber \\
&&+D^2\kappa_I^2\left( 1+\frac{\kappa_I^2}{q^2}  \right) (f_{IS}+f_{SI})+2D_I\kappa_I^2\varepsilon_0 \kB T \delta(t) 
\end{eqnarray}

\item Transverse components:
\begin{eqnarray}
\frac{1}{V}\langle \PP_T(\qq,t)\cdot\PP_T(-\qq,0)  \rangle &=&2D_S\kappa_S^2 \frac{\ex{-(D_S q^2+2D_S^r)t}}{D_S q^2+2D_S^r} \\
\frac{1}{V}\langle \PP_T(\qq,t)\cdot\jj_T(-\qq,0)  \rangle &=&\langle \jj_T(\qq,t)\cdot\PP_T(-\qq,0) = 0 \\
\frac{1}{V}\langle \jj_T(\qq,t)\cdot\jj_T(-\qq,0)  \rangle &=& 4D_I\kappa_I^2\varepsilon_0 \kB T \delta(t)
\end{eqnarray}

\end{itemize}

We can then deduce  the expressions of the longitudinal and transverse components of the susceptibility tensors [Eqs.~\eqref{suscep1}-\eqref{suscep4}], whose imaginary parts are plotted in the main text.

\section{Numerical parameters}

The numerical parameters used for the plots are given in Table~\ref{tab_num_params}.

\begingroup
\renewcommand{\arraystretch}{1.8}
\begin{table}[]
\begin{center}
\begin{tabular}{c|l|l} 
     \textbf{Symbol}&  \textbf{Definition} &\textbf{Value}\\ 
     \hline\hline
     $D_I$& 
 \makecell[l]{Diffusion coefficient of the ions, for a NaCl electrolyte \\ (average of $D_{\text{Na}^+}=\num{1.54e-9} \unit{m^2.s^{-1}}$ \\ and $D_{\text{Cl}^-}=\num{1.28e-9} \unit{m^2.s^{-1}}$)}& \num{2e-9} \unit{m^2.s^{-1}}\\
 \hline
 $D_S$& Diffusion coefficient of a solvent molecule&\num{2.3e-9} \unit{m^2.s^{-1}}\\
  \hline
  $D_S^r$& Rotational diffusion coefficient of a solvent molecule&\num{0.05} \unit{ps^{-1}}\\
\hline
$C_I$& Number density of the electrolyte&varies from $10^{-3}$ M to 1.2 M \\
 \hline
$a$& Typical `size' of a solvent molecule ($D_S /D_S^r = a^2$)&2.4 \AA\\
 \hline
$e$& Elementary charge&\num{1.6e-19} \unit{C}\\
 \hline
 $z$ & Valency of the electrolyte & 1 \\
  \hline
 $\varepsilon_0$& Permittivity of the vacuum&\num{8.9e-12} \unit{F.m^{-1}} \\
  \hline
  $\varepsilon_\text{w}$& Relative permittivity of pure water &78.5 \\
   \hline
 $\kB T$& Thermal energy at room temperature&\num{4e-21} \unit{J}\\
  \hline
 $\kappa = \sqrt{\frac{2C_0 z^2 e^2 }{\kB T \varepsilon}}$ & Inverse Debye length (electrolyte) &  varies from $0.01$ \AA$^{-1}$ to $0.36$ \AA$^{-1}$\\
  \hline
  $\kappa_{S}(0) = \sqrt{\frac{2C_S p^2 }{3\kB T \varepsilon a^2}}$ & Inverse Debye length of the solvent (at $q\to 0$) &  5.18 \AA$^{-1}$ \\
 \end{tabular}
\end{center}
    \caption{Parameters used in our numerical estimates.}
    \label{tab_num_params}
\end{table}
\endgroup


%